\documentclass[transmag]{IEEEtran}

\usepackage{latexsym}
\usepackage{graphicx}
\usepackage{amsfonts,amssymb,amsmath}
\usepackage[hidelinks]{hyperref}
\usepackage{acronym}
\usepackage{xcolor}
\usepackage[shortlabels]{enumitem}
\usepackage{subfigure}
\usepackage{cite}
\usepackage{comment}

\newcommand{\bfg}[1]{\boldsymbol{#1}}
\newcommand{\bfp}[1]{\boldsymbol{#1}'}
\newcommand{\bfpp}[1]{\boldsymbol{#1}''}
\newcommand{\bfppp}[1]{\boldsymbol{#1}'''}
\newcommand{\bfd}[1]{\dot{\boldsymbol{#1}}}
\newcommand{\bfdd}[1]{\ddot{\boldsymbol{#1}}}

\newcommand{\bfb}[1]{\boldsymbol{\rm #1}}

\newcommand{\jj}{\jmath}

\newcommand{\e}[1]{\boldsymbol{\rm e}_{\rm #1}}

\newcommand{\T}{\bfg T(t)}
\newcommand{\B}{\bfg B(t)}
\newcommand{\N}{\bfg N(t)}
\newcommand{\Td}{\dot{\bfg T}(t)}
\newcommand{\Bd}{\dot{\bfg B}(t)}
\newcommand{\Nd}{\dot{\bfg N}(t)}

\renewcommand{\th}[1]{\theta_{#1}}

\renewcommand{\sc}[2]{{\rm sc}(\th{#1}, \th{#2})}

\newcommand{\fourier}[1]{\mathfrak{#1} ( \omega )}
\newcommand{\hilbert}[1]{\hat{#1} ( t )}
\newcommand{\analytic}[1]{\tilde{#1} ( t )}
\newcommand{\FT}[1]{\mathcal{F}\left [ #1 \right ]( \omega )}
\newcommand{\HT}[1]{\mathcal{H}\hspace{-0.4mm}\left [ #1 \right ]\hspace{-0.9mm}( t )}
\newcommand{\AS}[1]{#1 ( t ) + \jj \hilbert{#1}}
\newcommand{\interval}{( -\infty, \infty )}
\newcommand{\ampli}[1]{{\rm #1}}
\renewcommand{\exp}[2]{#1 \, e^{\jj \, #2}}
\newcommand{\wk}{\omega_{\kappa}(t)}
\newcommand{\wt}{\omega_{\tau}(t)}

\acrodef{rocof}[RoCoF]{Rate of Change of Frequency}

\acrodef{if}[IF]{Instantaneous Frequency}
\acrodef{ht}[HT]{Hilbert Transform}
\acrodef{ft}[FT]{Fourier Transform}
\acrodef{ha}[HA]{Harmonic Analysis}
\acrodef{ct}[CT]{Clarke Transform}
\acrodef{as}[AS]{Analytic Signal}
\acrodef{pll}[PLL]{Phase-Locked Loop}

\def\BibTeX{{\rm B\kern-.05em{\sc i\kern-.025em b}\kern-.08em T\kern-.1667em\lower.7ex\hbox{E}\kern-.125emX}}

\begin{document}

\title{Using Differential Geometry to Revisit the Paradoxes of the Instantaneous Frequency}

\author{Federico~Milano,~\IEEEmembership{Fellow,~IEEE}, Georgios
  Tzounas,~\IEEEmembership{Member,~IEEE}, \\ Ioannis Dassios, Mohammed Ahsan Adib Murad, and
  Taulant K\"{e}r\c{c}i, \IEEEmembership{Member,~IEEE}%
  \thanks{F.~Milano, G.~Tzounas, I.~Dassios are with School of Electrical and Electronic Engineering, University College Dublin, Dublin, D04V1W8, Ireland.  e-mails:
    \{federico.milano, georgios.tzounas, ioannis.dassios\}@ucd.ie}%
  \thanks{M.~A.~A.~Murad is with DIgSILENT GmbH, Germany.}%
  \thanks{T.~K\"{e}r\c{c}i is with EirGrid plc, Ireland.}%
  \thanks{This work is supported by the Sustainable Energy Authority of Ireland~(SEAI), by funding F.~Milano and I.~Dassios under the project FRESLIPS, Grant No.~RDD/00681; and by the European Commission, by funding G.~Tzounas and F.~Milano under the project EdgeFLEX, Grant No.~883710.}%
}

\IEEEtitleabstractindextext{%

  \begin{abstract}
    This paper proposes a general framework to interpret the concept
    of \acf{if} in three-phase systems.  The paper first recalls the
    conventional frequency-domain analysis based on the Fourier
    transform as well as the definition of \ac{if} which is based on
    the concept of analytic signals.  The link between analytic
    signals and Clarke transform of three-phase voltages of an ac
    power system is also shown. Then the well-known five paradoxes of
    the \ac{if} are stated.  In the second part of the paper, an
    approach based on a geometric interpretation of the frequency is
    proposed.  This approach serves to revisit the five \ac{if}
    paradoxes and explain them through a common framework.  The case
    study illustrates the features of the proposed framework based on
    a variety of examples and on a detailed model of the IEEE 39-bus
    system.
  \end{abstract}


}

\maketitle

\section{Introduction}
\label{sec:intro}

\subsection{Motivation}

The recent move toward non-synchronous and distributed generation has led the power system community to rediscuss not only the operation and the economic aspects of the electric grid but also its very basic mathematical foundation \cite{9286772, 8450880, Paolone:2020}.  
This paper focuses on a fundamental quantity of ac grids, namely the frequency, its meaning and its many definitions, as well as the paradoxes that these definitions appear to generate.  In this context, the paper aims at proposing a common framework, based on the concept of \textit{invariant} provided by differential geometry, that can explain these paradoxes.

\subsection{Literature Review}

The \acf{if} of an ac signal is defined as the time derivative of the signal's phase angle \cite{IEEE118}.  This definition assumes that the signal is described by a single sinusoid.  If the representation of the signal is a different function, for example the sum of two sinusoids, then the definition of its frequency is less straightforward.  As a matter of fact, the literature on this topic is vast, as much as the different signal representations that have been considered \cite{8586583}.

Another issue that has been widely discussed in the literature is that the value of frequency seems to depend on the transformation utilized to represent the signal itself. This is an apparent inconsistency since, intuitively, 
the estimated frequency should be the same independently of the transformation.  This issue is clearly relevant in engineering applications, for example in the design of control systems that regulate the frequency in a power system.  If the estimation of the controlled signal is not correct or accurate enough, then it also becomes hard to ensure a reliable and robust control design \cite{6704843, 6939460, 8378935, 9494963, TAN2022107362}.  

Yet another issue with the common definition of frequency and a large number of existing frequency estimation techniques is that they do not account for variations of the magnitude of the signal.  However, the magnitude of measured signals, e.g., the voltage at a network bus, is not constant during electromechanical transients, which is when it is of upmost importance to be able to accurately estimate frequency variations \cite{phadke2008synchronized, FDF:2020}.  Thus, only techniques that are able to measure the phase angle independently of the magnitude of the measured signals are useful in power system applications.  

A last issue is that the estimation of the \ac{if} should be available \textit{as soon as possible}.  This requirement conflicts with the need of certain transform-based techniques, such as the \acf{ft} and the \acf{ht}, for several samples -- in principle, infinitely many -- of the signal \cite{Cohen:1995,  Hahn:1996}.  Also in this case, several techniques that involve, e.g., mobile windows, anti-aliasing, etc., have been proposed \cite{roscoe2013p, bertocco2015compressive, 8928939}, although the need for a minimum set of measurements poses intrinsic and probably unsolvable limits to this kind of approaches \cite{8675542}.  The inability of the \ac{ft} to track the \ac{if} is what has led, ultimately, to the well-known paradoxes that are discussed in this work.  

An option to overcome the limitations of the \ac{ft} is to use functions other than sine and cosine waves to decompose the signal.  This has been done for example with wavelets, which include both a frequency and a damping, and thus appear more adequate to represent the profile of electromechanical oscillations observed in power systems \cite{6595731, mei2006dynamic, turunen2010selecting}.  The ultimate resource is to use heuristic custom functions to decompose the signal, which is, in turn, the solution provided by the Hilbert-Huang Transform~(HHT) \cite{Huang:2014}.  For its flexibility, the HHT has found applications also in the analysis of power system transients \cite{1664959, 4813184, 8338416}.  However, this kind of adaptive transforms do not solve the problem of the estimation of the \ac{if} \cite{8467370}.

Another family of techniques, which are based on \acfp{pll}, attempt to \textit{track} the signal while it is evolving in time \cite{karimi2004estimation, nicastri2010comparison}.  The estimation of the frequency based on \acp{pll} is conceptually closer to the definition of \ac{if} given in \cite{IEEE118}.  However, \acp{pll} rely on a control loop, which can be also quite sophisticated, but ultimately suffers from another intrinsic dilemma: the faster is the tracking the more sensitive to noise is the estimation.  

A recent interpretation of electric quantities as ``curves'' suggests that the frequency corresponds to the \textit{curvature} of a trajectory \cite{freqgeom}.  This interpretation is appealing as the curvature is a geometric invariant and, as such, is independent from the coordinates that are used to define the signal. However, the analogy also comes with some unexpected byproducts, e.g., the fact that, if the curve has three (or more) dimensions, then there exist more than one invariant (and, hence more than one frequency) that define the trajectory \cite{freqfrenet}.  

\subsection{Contributions}

The contributions of this work are twofold.  First, it provides an overview of the 
existing transforms that have been traditionally utilized to define the ``frequency'' of a signal.  This overview prepares the ground for presenting the paradoxes of \ac{if}.  While the focus of this overview is on \ac{ft}, \ac{ht} and analytic signals, we also discuss the links of these techniques with quantities and techniques employed in circuit theory and power system analysis, such as phasors, \acf{ha} and \acf{ct}.  The second contribution is to show that conventional techniques can be revisited in terms of a geometric approach.  This approach is used as a general framework where \ac{ft}, \ac{ht} and analytic signals (and hence also phasors, \ac{ha} and \ac{ct}) can be interpreted in terms of special systems of coordinates and curves.  The paradoxes of the \ac{if} are then revisited using this framework.

\subsection{Organization}

The remainder of this paper is organized as follows.  Section~\ref{sec:background} presents conventional techniques for the definition of the frequency of a signal.  This section also presents the classical paradoxes of the \ac{if}.  Section~\ref{sec:geometry} introduces the geometric approach that is utilized in the paper to revisit conventional techniques and the paradoxes.  Section~\ref{sec:examples} discusses the proposed approach by means of analytic examples and a case study based on an EMT model of the IEEE 39-bus system. Section~\ref{sec:conclusions} draws relevant conclusions.

\section{Background on Frequency and Time Analysis}
\label{sec:background}

Electrical quantities such as the voltage and the current can be expressed, as many other quantities in physics and engineering, as functions of time $t$.  For example, a steady-state ac voltage can be represented in time domain as:
\begin{equation}
  \label{eq:v}
  v(t) = V \cos ( \omega_o t ) \, ,
\end{equation}
where $V$ and $\omega_o$ are the amplitude and the angular frequency, respectively, of the voltage.

In signal processing, ``signals'' are also often called ``time waveforms'', which stresses the attention on the wave-like and thus potentially periodic nature of signals.  However, in the remainder of this work we refrain from considering that signals are necessarily periodic.  On the contrary, we focus precisely on the cases for which signals undergo a transient.  Using the notation of \eqref{eq:v}, one has:
\begin{equation}
  \label{eq:vt}
  v(t) =  V(t) \cos \big (\vartheta(t) \big ) \, ,
\end{equation}
where 
$V(t)$ and 
$\vartheta(t)$ are arbitrary functions of time.

\subsection{\acf{ft}}
\label{sub:fourier}

The \ac{ft} (or \textit{spectrum}) of the signal $v(t)$ is defined as: 
\begin{equation}
  \label{eq:fourier}
  \fourier{v} = \FT{v} = \frac{1}{\sqrt{2\pi}} \int v(t) e^{-\jj \omega t} dt \, , 
\end{equation}
where $\jj$ is the imaginary unit and the integral has to be intended to be calculated in the range $t \in \interval$.  The function $\fourier{v}$ provides a representation of the signal $v(t)$ in the frequency domain or space.  The fact that the frequency domain is, in effect, an alternative \textit{space} with its own coordinate, has a relevant role in the discussion given in this paper and is thus a point that we wish to highlight here.

The \ac{ft} works the best, and in fact it was invented specifically for, stationary periodic signals.  For \eqref{eq:v}, one obtains:
\begin{equation}
  \label{eq:ft:v}
  \begin{aligned}
    \fourier{v}
    &= \FT{V \cos (\omega_o t))} \\
    &= \sqrt{2\pi} \, \tfrac{V}{2} \,
    \big (\delta(\omega - \omega_o) + \delta(\omega + \omega_o) \big ) \, ,
  \end{aligned}
\end{equation}
which is a spectrum with non-null values only in $-\omega_o$ and $\omega_o$, as the time-domain signal has only one frequency.  The idea of the \ac{ft} is that, if a signal can be expressed (or approximated) with a series of sinusoids, then one obtains a sharp spectrum with only values corresponding to the frequencies of the sinusoids that compose the time-domain signal.

The application of the \ac{ft} to power system analysis has found its natural field in the \ac{ha}, i.e. the study of the effect of frequencies multiple of the fundamental one in stationary conditions, e.g.,~see \cite{das}.  In a \ac{ha}, a signal can be represented as a sum of sinusoids and, possibly, a non-null constant term:
\begin{equation}
  \label{eq:vh}
  v(t) = \sum_{h} V_h \cos(h \omega_o t + \theta_h)
  \, , \quad h \in \{0, 1, \dots, n\} \, ,
\end{equation}
which can be conveniently studied in the frequency domain $\{ 0, \omega_o, \dots, h\omega_o, \dots, n\omega_o \}$ rather than in the time domain.

Difficulties arise, however, when the signal is not periodic.  The spectrum becomes a continuum rather than a set of sharp frequency values.  Moreover, the integral of \eqref{eq:fourier} has to be calculated for $t \in \interval$, which is impractical for the vast majority of real-world applications.   This issue is particularly relevant if one wants to apply the (discrete) \ac{ft} to a measured voltage that evolves during an electromechanical transient.

Several \textit{patches}, more or less sophisticated, have been proposed to compensate the inevitable approximations required to calculate the \ac{ft} of a non-stationary signal.  The most used approach is the short-time discrete Fourier transform (sDFT).  This utilises a ``windowed'' signal, i.e.~the actual signal is multiplied by a function that is nonnull only in the interval of time of interest for the estimation of the frequency.  While the sDFT makes possible the calculation of the Fourier transform as it restricts the integration of the signal to a finite interval, it also introduces, as it is to be expected, some issues, such as aliasing and spectral leakage.  The literature provides plenty of techniques to reduce the impact of these issues on the estimation.  Among these, the appropriate choice of the window profile and length, sampling rate, and spectral interpolation.  The literature on this topic is vast.  We refer the reader to the review of most common techniques provided in Chapter 3 of \cite{Milano:2017}.   Relevant recent works are, e.g., \cite{8675542} and \cite{8630817}.   The main conclusion that can be drawn from existing literature is that the \ac{ft} can be adapted to transient signals and provide a relatively good estimation of the frequency variations.  It remains, however, the fundamental issue that the \ac{ft} is meant for stationary periodic signals.

\subsection{\acf{ht} and Analytic Signals}
\label{sub:hilbert}

For the definition of the \acf{if} it is convenient to define a mathematical object called \textit{analytic signal}.  This is a complex quantity, which is calculated from the signal $v(t)$ as follows:
\begin{equation}
  \label{eq:as0}
  \analytic{v} = v(t) + \frac{\jj}{\pi} \int \frac{v(r)}{t-r} dr \, . 
\end{equation}
This simplified notation is also utilized in the remainder of this paper.  The imaginary part of $\analytic{v}$ is the \ac{ht} of $v(t)$ \cite{Hahn:1996}:
\begin{equation}
  \HT{v} = \hilbert{v} =
  \frac{1}{\pi} \int \frac{v(r)}{t-r} dr \, .
\end{equation}
The analytic signal $\analytic{v}$ can thus be written equivalently as:
\begin{equation}
  \label{eq:as}
  \analytic{v} = v(t) + \jj \, \HT{v} = \AS{v} \, .
\end{equation}

Differently from most transforms utilized in signal processing, the \ac{ht} retains the domain of the signal and, in fact, returns a function of time.  It is relevant to observe since now that the \ac{ht} is often interpreted as a \textit{rotation} of $-\pi/2$ of the signal to which it is applied.  This notion is justified from the fact that the \ac{ht} of the sine and cosine functions are:
\begin{align}
  \mathcal{H}[\cos](\omega t) &=  \sin(\omega t) \, , \\
  \mathcal{H}[\sin](\omega t) &= -\cos(\omega t) \, ,
\end{align}
for $\omega > 0$.  For negative frequencies, the signs of the right-hand side of the equalities above are swapped.  To avoid this issue, analytic signals are conventionally defined only for $\omega > 0$.  The rotation can be formalized observing that:
\begin{equation}
  \label{eq:ft:ht}
  \FT{\hat{v}} = - \jj \, \FT{v} \, , \qquad \omega > 0 \, .
\end{equation}
While it is intuitive to appreciate a rotation of a periodic signal, less clear is the meaning of a rotation for an arbitrary signal.

It is also relevant to note that the effect of the \ac{ht} has a resemblance with that of the time derivative of the signal:
\begin{equation}
  \label{eq:ft:dt}
  \FT{v'} = \jj \, \omega \, \FT{v} \, , 
\end{equation}
where $v'(t) = \frac{d v(t)}{dt}$.  Merging \eqref{eq:ft:ht} and \eqref{eq:ft:dt} gives:
\begin{equation}
  \label{eq:ft:ht:dt0}
  \FT{\hat{v}} = - {\FT{v'}}\big/{\omega} \, ,
\end{equation}
and, hence:
\begin{equation}
  \label{eq:ft:ht:dt}
  \hat{v}(t) = - \mathcal{F}^{-1} \left [ {\FT{v'}} \big/{\omega} \right ](t) \, ,
\end{equation}
which shows how the \ac{ht} of a signal is related to the \ac{ft} of the signal itself as well as the way to calculate it.

The Bedrosian theorem provides an important property of the \ac{ht}. This theorem proves that if two functions $\analytic{f}$ and $\analytic{g}$ are analytic and if $\fourier{f} = \FT{\tilde{f}}$ vanishes for $|\omega| > a$ and $\fourier{g} = \FT{\tilde{g}}$ vanishes for $|\omega| < a$, where $a$ is a positive constant, then the following identity holds:
\begin{equation}
  \label{eq:bi}
  \HT{\tilde{f} \, \tilde{g}} = \analytic{f} \, \HT{\tilde{g}} \, .
\end{equation}

The Bedrosian identity \eqref{eq:bi} has a special role in power system analysis and the estimation of frequency variations during electromechanical transients.  In fact, in electromechanical transients, the time-varying amplitude of the voltage has a low-frequency spectrum that does not overlap the high-frequency spectrum of the phase of the voltage itself \cite{8928939}.  In turn, thus, 
if
the voltage is expressed by \eqref{eq:vt} and is undergoing an electromechanical transient, 
applying the \ac{ht} to $v(t)$ gives:
%
\begin{equation}
  \label{eq:biv}
  \begin{aligned}
    \HT{v}
    &= \HT{V \, \cos(\vartheta)} 
    = V(t) \, \HT{\cos(\vartheta)} \\
    &= V(t) \, \sin(\vartheta(t)) \, .
  \end{aligned}
\end{equation}
Hence, the analytic signal of the voltage can be written as:
\begin{equation}
  \label{eq:as2}
  \begin{aligned}
    \analytic{v}
    &= V(t) \, \left ( \cos(\vartheta(t)) + \jj \sin(\vartheta(t)) \right ) 
    = \exp{V(t)}{\vartheta(t)} \, ,
  \end{aligned}
\end{equation}
and, hence, its \ac{if} can, in theory, be calculated without having to know $V(t)$.  We say \textit{in theory} because one has to be able to calculate $\HT{v}$, which is obtained as an integral for $t \in \interval$.  This issue is further discussed in Section \ref{sub:five1} and constitutes, in effect, one of the five paradoxes of the \ac{if}.  Another issue is that, in practice, the (discrete) \ac{ht} is in fact calculated using the (discrete) \ac{ft} and its inverse as indicated by \eqref{eq:ft:ht:dt}.  Thus, apart from its intrinsic issues, the utilization of the \ac{ht} also suffers of the issues of the \ac{ft} discussed in Section \ref{sub:fourier}.

The notation of \eqref{eq:as2} is well-known in the analysis of ac circuits and power systems, where, for historical reasons, is called \textit{phasor} and is generally utilized in stationary conditions and shifted by the fundamental frequency $\omega_o$.  In ac circuit analysis, the \ac{ht} and analytic signals are not well-known nor needed, as a matter of fact, for the definition of phasors.  This is because phasors, by definition, are characterized by a unique frequency, say $\omega_o$.  On the other hand and for the same reason, the phasors of a circuit require to be referred to a common reference phase angle, say $\theta_o$, which defines unequivocally yet arbitrarily the coordinates -- rotating with angular speed $\omega_o$ -- with respect to which the real and imaginary parts of the phasors
are defined.  In turn, thus, phasors are analytic signals shifted by $\omega_o t + \theta_o$.



Finally, it is relevant to observe that, given the linearity of the \ac{ht}, applying it to the signal defined in \eqref{eq:vh} gives:
\begin{equation}
  \label{eq:ht:vh}
  \begin{aligned}
    \HT{\sum_h V_h \cos(h \omega_o t + \theta_h)} &=
    \sum_h \HT{V_h \cos(h \omega_o t + \theta_h)}  \\
    &=\sum_h V_h \sin(h \omega_o t + \theta_h) \, ,
  \end{aligned}
\end{equation}
and, hence, one can define the analytic signal of a sum of sinusoids as
the sum of the analytic signals of 
these sinusoids.  
Thus, the analytic
signal associated with each harmonic $h$ is:
\begin{equation}
  \label{eq:as:vh}
  \tilde{v}_h(t) =
  V_h \cos(h \omega_o t + \theta_h) +
  \jj V_h \sin(h \omega_o t + \theta_h) \, . 
\end{equation}

\subsection{\acf{ct}}
\label{sub:clarke}

We have considered so far exclusively an individual ``signal,'' which can be opportunely manipulated with time- and frequency-domain transforms.  Power systems, however, are mostly three-phase circuits.  The voltage of a node is thus a triplet of measurements rather than a single quantity.  Of course, one can treat the voltage of each phase 
as a signal and proceed with the analysis discussed so far.  But having more then one phase provides additional information.

The framework proposed in this paper extends the notion of a signal to that of a ``curve'' and, as such, it assumes a multi-dimensional space.  Limiting our analysis to three dimensions, the phases of an ac three-phase system appear as ideal candidates for the definition of such a space.  Under certain conditions, however, the dimension of this space can be reduced to two.  This is, in turn, the goal of the \ac{ct}. 

Let $\bfg v_{abc}(t) = (v_a(t), v_b(t), v_c(t))$ be the voltage triplet of a three-phase node.  The \ac{ct} applied to this signal returns another triplet, say $\bfg v_{\alpha \beta \gamma}(t) = (v_{\alpha}(t), v_{\beta}(t), v_{\gamma}(t))$ calculated as:
\begin{equation}
  \label{eq:clarke}
  \begin{aligned}
    \bfg v_{\alpha \beta \gamma }(t)
    &= \bfb C \, \bfg v_{abc}(t) \\
    &= \frac{2}{3}
    \begin{bmatrix}
      1 & -\frac{1}{2} & -\frac{1}{2} \\
      0 &  \frac{\sqrt{3}}{2} & -\frac{\sqrt{3}}{2} \\
      \frac{1}{\sqrt{2}} & \frac{1}{\sqrt{2}} & \frac{1}{\sqrt{2}} \\
    \end{bmatrix}
    \begin{bmatrix}
      v_{a}(t) \\
      v_{b}(t) \\
      v_{c}(t)
    \end{bmatrix} \, .
  \end{aligned}
\end{equation}

\begin{figure}[!ht]
  \begin{center}
    \subfigure[Space $(v_a, v_b, v_c)$]{\resizebox{0.475\linewidth}{!}{\includegraphics{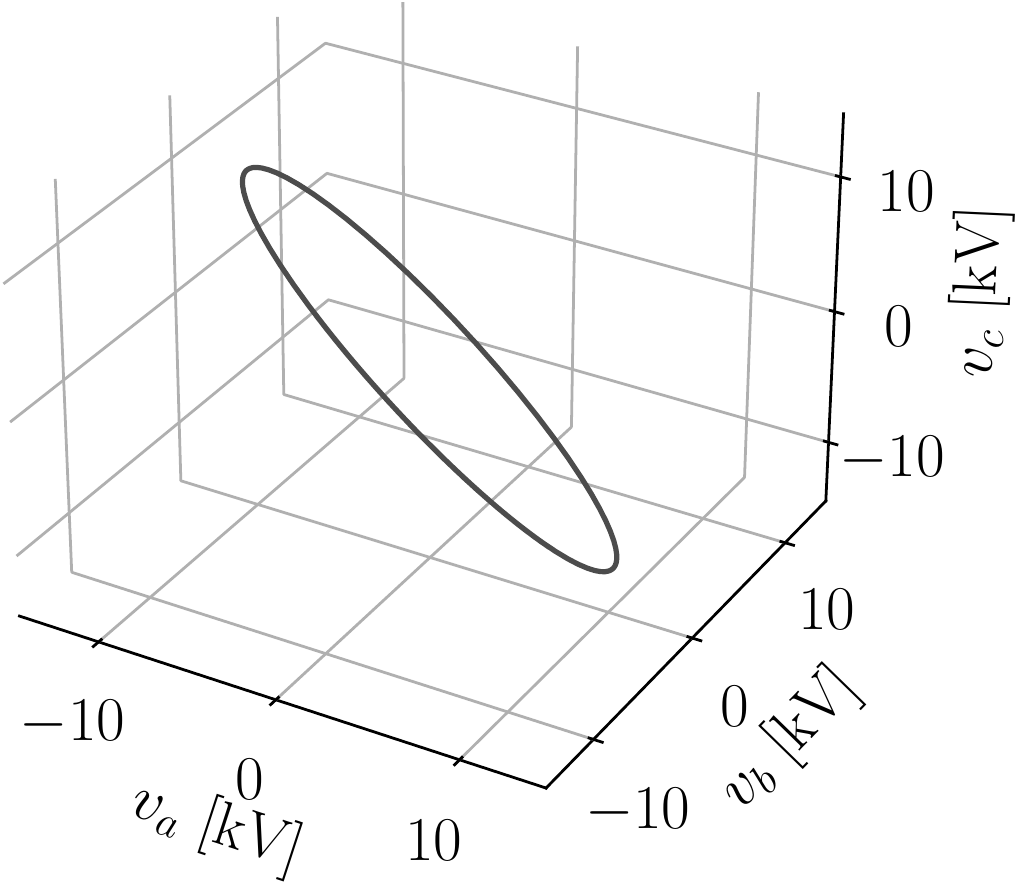}}}
    \subfigure[Plane $(v_{\alpha}, v_{\beta})$]{\resizebox{0.475\linewidth}{!}{\includegraphics{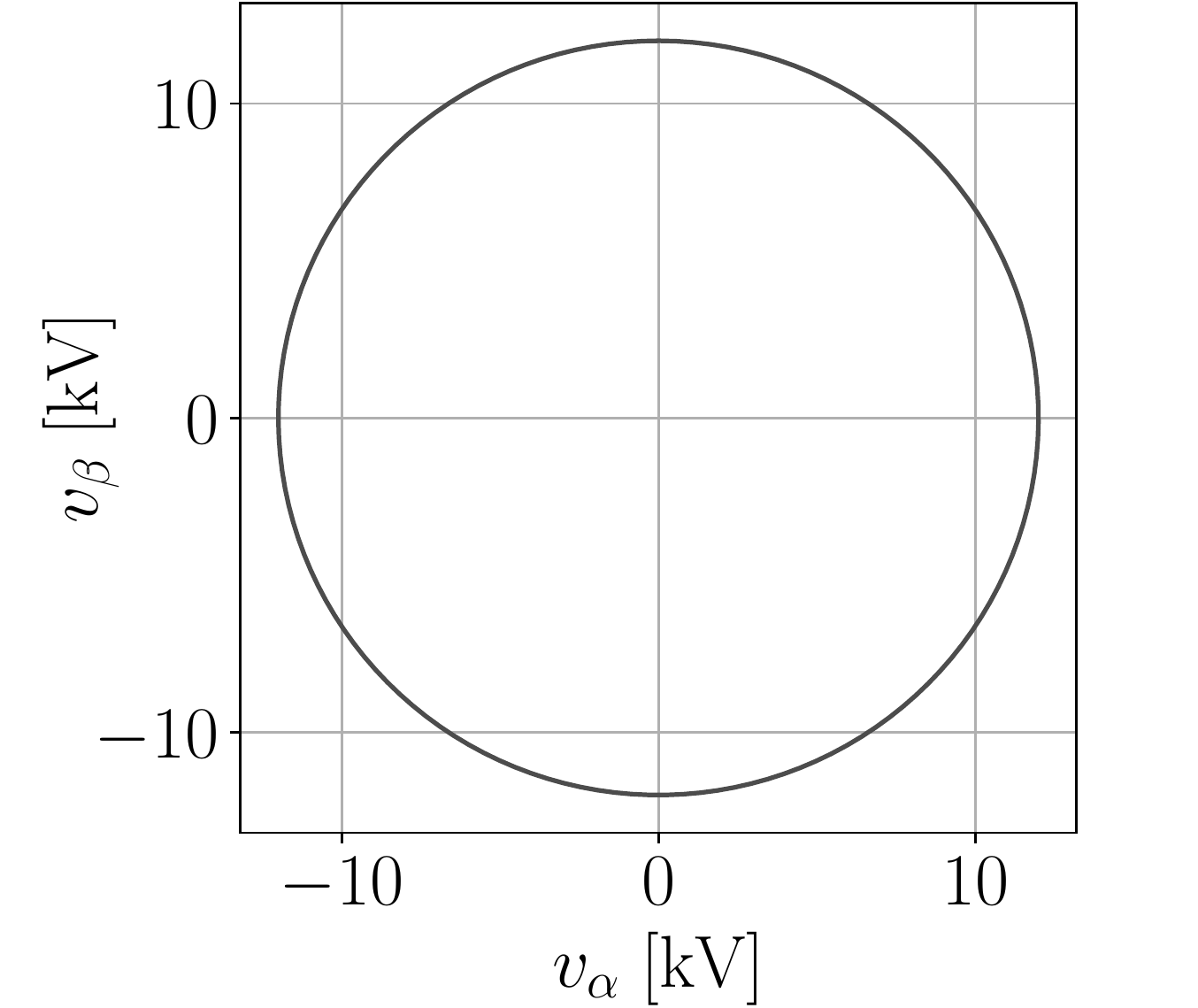}}}
    \caption{Balanced positive-sequence
      three-phase voltage with $V = 12$ kV.}
    \label{fig:clarke}
  \end{center}
\end{figure}

In general, the three components of $\bfg v_{\alpha \beta \gamma}(t)$ are non-null.  However, in the special case of balanced voltages, namely:
\begin{equation}
  \begin{aligned}
    v_a(t) &= V(t) \cos \big ( \vartheta (t) \big ) \, , \\
    v_b(t) &= V(t) \cos \big ( \vartheta (t) - \tfrac{2}{3} \pi \big ) \, , \\
    v_c(t) &= V(t) \cos \big ( \vartheta (t) + \tfrac{2}{3} \pi \big ) \, ,
  \end{aligned}
\end{equation}
the \ac{ct} gives:
\begin{equation}
  \label{eq:ct}
  \begin{aligned}
    v_{\alpha}(t) &= V(t) \cos \big ( \vartheta (t) \big ) \, , \\
    v_{\beta}(t) &= V(t) \sin \big ( \vartheta (t) \big ) \, , \\
    v_{\gamma}(t) &= 0 \, .
  \end{aligned}
\end{equation}
The latter expression can be rewritten as a complex quantity:
\begin{equation}
  \label{eq:ct2}
  \bar{v}(t) = v_{\alpha}(t) + \jj \, v_{\beta}(t) \, ,
\end{equation}
which has a striking resemblance with an analytic signal.  This can be readily observed in Fig.~\ref{fig:clarke} that shows a balanced positive-sequence three-phase voltage.  The curve generated by the voltage $\bfg v_{abc}(t)$ is a circle that lies in the plane $(v_{\alpha}, v_{\beta})$.  Reference \cite{freqfrenet} shows that imbalances in the amplitude and/or phase angles of the voltage also lead to a plane curve, typically an ellipse rather than a circle.  However the plane of such curve is not the plane defined by the
$\alpha \beta$-frame.

The quantity defined in \eqref{eq:ct2} is not, in general, an analytic signal as it can contain negative frequencies.  The simplest case for which this happens is the stationary conditions for which the voltage $\bfg v_{\alpha \beta \gamma}(t)$ contains a positive and a negative sequence.  Then, applying the symmetrical component transform and observing that the \ac{ct} is a linear operator, one obtains:
\begin{equation}
  \label{eq:ct3}
  \begin{aligned}
    \bar{v}(t) &= \bar{v}^+(t) + \bar{v}^-(t) \\
    &= \big ( v^+_{\alpha}(t) + \jj \, v^+_{\beta}(t) \big ) +
    \big ( v^-_{\alpha}(t) + \jj \, v^-_{\beta}(t) \big ) \, ,
  \end{aligned}
\end{equation}
where $\bar{v}^+(t)$ has angular frequency $\omega_o$ and $\bar{v}^-(t)$ has angular frequency $-\omega_o$.  Clearly, $\bar{v}^+(t)$ is an analytic signal.  At this point, one might raise the question why analytic signals are defined only for positive frequencies.  This is a consequence of the fact that they are obtained starting from a single signal, not a triplet.  This leads to have no way to define the sign of the frequency itself.  In fact, the spectrum of a signal is symmetrical and one can use conveniently only the part for $\omega > 0$, which is the common choice for analytic signals.

Finally, we note that applying the Park transform to $\bar{v}(t)$ in \eqref{eq:ct2} leads to the Park vector \cite{freqcomplex}.  According to the discussion above, the Park vector, like more conventional phasors, is just $\bar{v}(t)$ shifted by $\omega_{\rm p} t + \theta_{\rm p}$, where $\omega_{\rm p}$ and $\theta_{\rm p}$ are the angular speed and the phase reference of the Park $\rm dq$-axis rotating frame.  From this definition, it descends that, in stationary conditions and for $\omega_{\rm p} = \omega_o$, the Park vector ``downgrades'' to a phasor.  Equivalently, phasors can be seen as steady-state Park vectors.

\subsection{\acf{if}}
\label{sub:if}

The importance of the \ac{ht} is largely due to the fact that it allows determining the \ac{if} of a time-varying signal.  We note that, since an analytic signal is a complex quantity, one has:
\begin{equation}
  \label{eq:as3}
  \analytic{v} = \exp{\ampli{v}(t)}{\phi(t)} \, ,
\end{equation}
where
\begin{equation}
  \label{eq:as4}
  \begin{aligned}
    \ampli{v}(t) &= |\analytic{v}| = \sqrt{v^2(t) + \hat{v}^2(t)} \, , \\
    \phi(t) &= \angle \analytic{v} = {\rm arctan} \left ( \frac{\hilbert{v}}{v(t)} \right ) \, .
  \end{aligned}
\end{equation}
Then, the \ac{if} is defined as:
\begin{equation}
  \label{eq:if}
  \phi'(t) = \frac{v(t) \hat{v}'(t) - \hilbert{v} v'(t)}{\ampli{v}^2(t)} \, .
\end{equation}

This definition has a very clear meaning only for signals in the form of \eqref{eq:vt}, for which $\ampli{v}(t) = V(t)$ and $\phi(t) = \vartheta(t)$, and, consequently, $\vartheta'(t)$ coincides with the \ac{if}.  However, in the most general case, $\phi(t)$ is not simply the phase of the original signal.  Perhaps the simplest example that shows this underlying complexity is the case of a signal composed of two harmonics:
\begin{equation}
  \label{eq:v12}
  v(t) = V_1 \cos(\omega_1 t) + V_2 \cos(\omega_2 t) \, ,
\end{equation}
with $\omega_1 > 0$ and $\omega_2 > 0$, which leads to the following analytic signal:
\begin{equation}
  \analytic{v} = \exp{V_1}{\omega_1 t} + \exp{V_2}{\omega_2 t} \, ,
\end{equation}
and to the following expression of the \ac{if} \cite{Cohen:1995}:
\begin{equation}
  \label{eq:f12}
  \phi'(t) = \frac{1}{2} (\omega_2 + \omega_1) +
  \frac{1}{2} \, \Delta \omega \, \frac{V^2_2 - V^2_1}{\ampli{v}^2(t)} \, ,
\end{equation}
where $\Delta \omega = \omega_2 - \omega_1$, and:
\begin{equation}
  \ampli{v}^2(t) = V_1^2 + V_2^2 + 2 V_1 V_2 \cos (\Delta \omega \; t) \, .
\end{equation}

\begin{figure}[!ht]
  \begin{center}
    \subfigure[Signal]{\resizebox{0.475\linewidth}{!}{\includegraphics{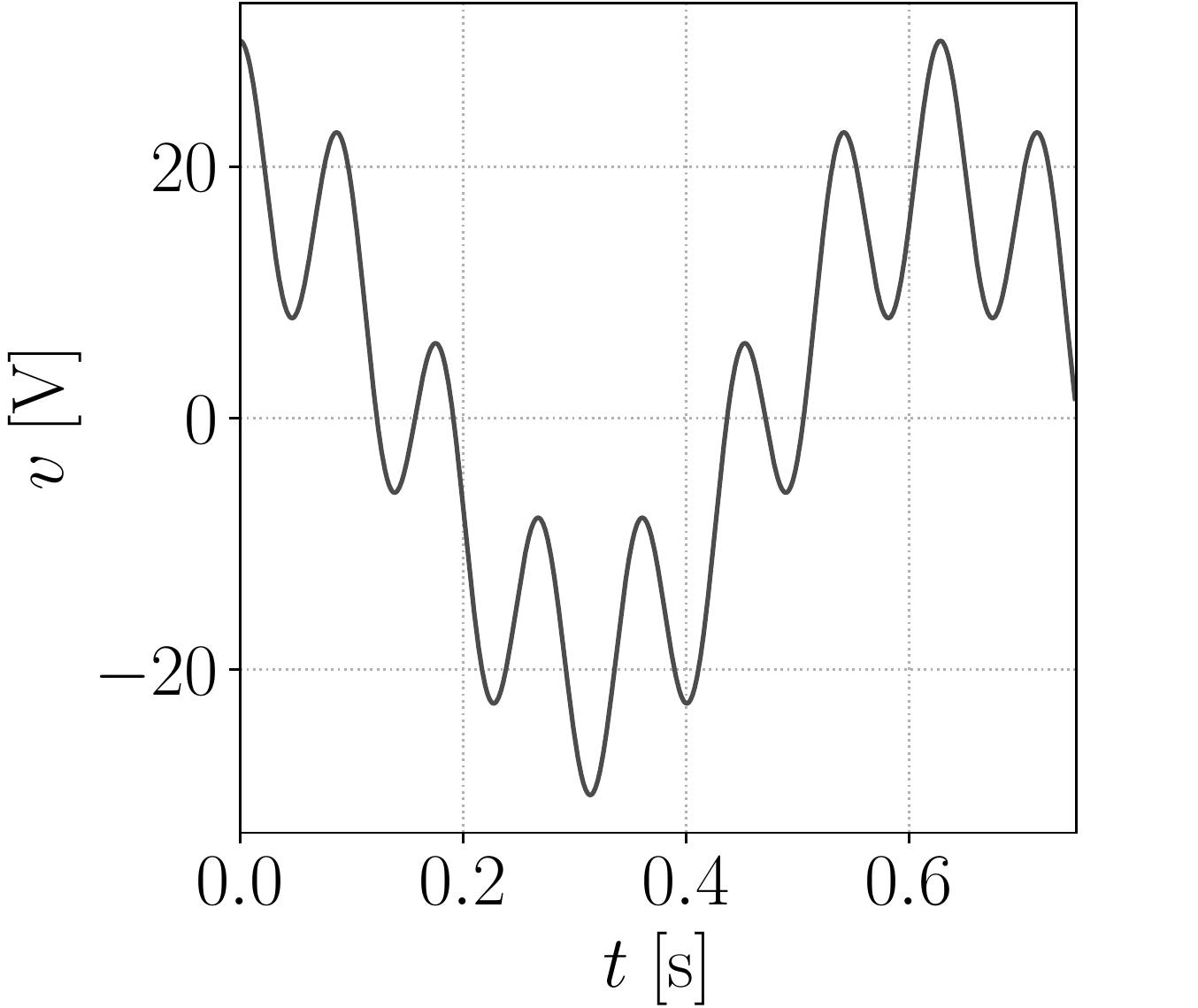}}}
    \subfigure[Instantaneous frequency]{\resizebox{0.475\linewidth}{!}{\includegraphics{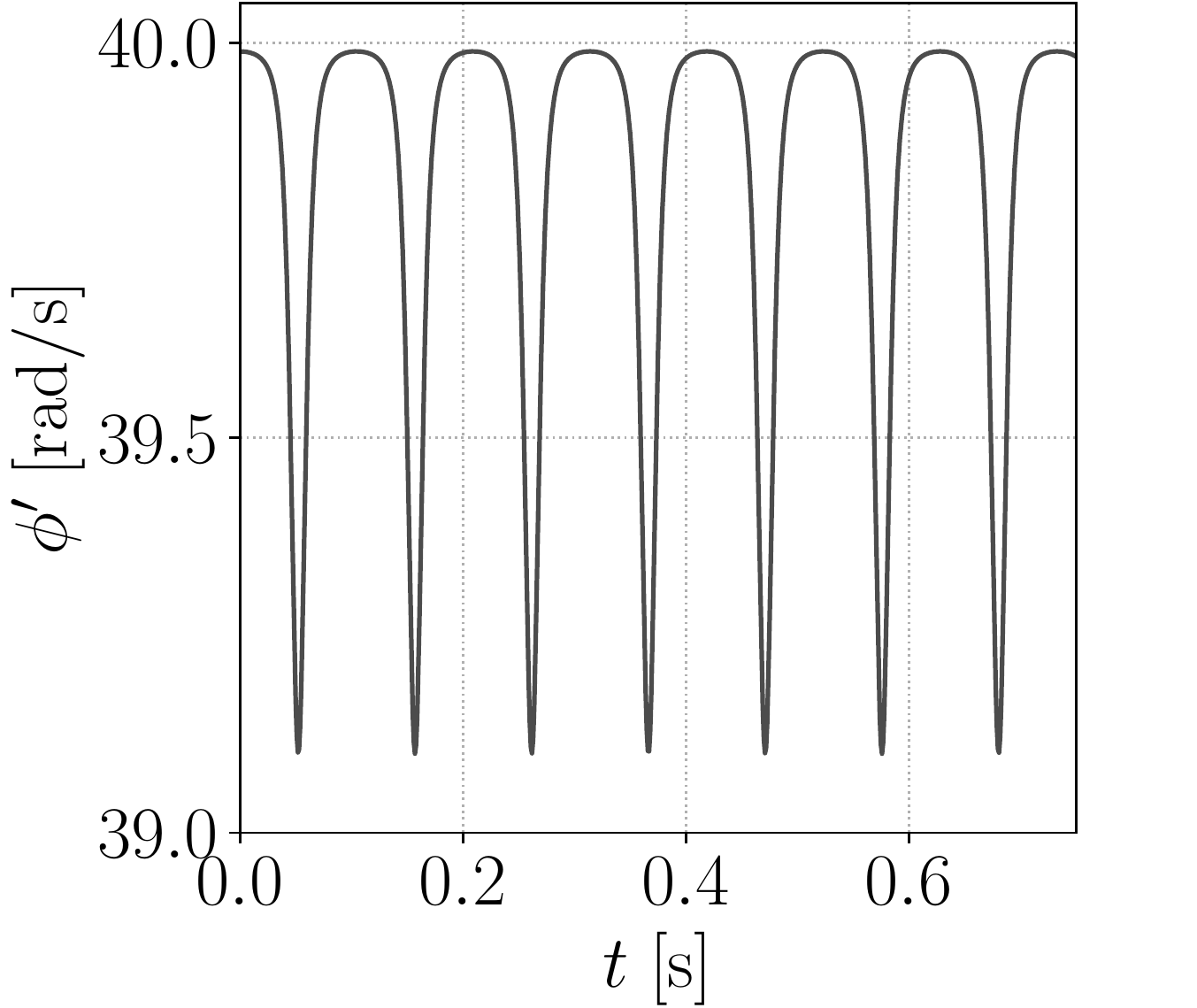}}}
    \caption{Representation of signal \eqref{eq:v12} and its \ac{if}.}
    \label{fig:v12}
  \end{center}
\end{figure}

Equation \eqref{eq:f12} shows that the \ac{if} of the signal \eqref{eq:v12} is, in general, not equal to $\omega_1$ and $\omega_2$ and, in fact, is not even constant if $V_1 \ne \pm V_2$.  Figure~\ref{fig:v12} illustrates the results obtained above assuming $\omega_1 = 10$, $\omega_2 = 70$, $V_1 = 20$, $V_2 = 10$.  The shape of the \ac{if} is somewhat surprising as the intuition  would lead to think that the \ac{if} of a signal the positive spectrum of which contains only two frequencies, namely $\omega_1$ and $\omega_2$ should be a simpler expression or, at least, a constant value.

\subsection{The Five Paradoxes of Instantaneous Frequency}
\label{sub:five1}

We are ready to present the paradoxes of the \ac{if} mentioned in the title of the paper.  In Chapter~2 of the book ``Time-frequency analysis,'' Leon Cohen presents five paradoxes on the \ac{if} \cite{Cohen:1995}.  These paradoxes are well-known in the 
area
of signal processing.  The time-frequency analysis is in effect an attempt to overcome the apparently irreconcilable differences between time and frequency domains by merging them into a unique framework.

We report verbatim below the five paradoxes as written by Cohen in \cite{Cohen:1995}.

\begin{enumerate}[label={\textbf{P\arabic*}:}]

\item \textit{\ac{if} may not be one of the frequencies in the spectrum.}

\item \textit{If we have a line spectrum consisting of only a few sharp frequencies, then the \ac{if} may be continuous and range over an infinite number of values.}

\item \textit{Although the spectrum of the analytic signal is zero for negative frequencies, the \ac{if} may be negative.}

\item \textit{For a band-limited signal the \ac{if} may go outside the band.}

\item \textit{If the \ac{if} is an indication of the frequencies that exists at time $t$, one would presume that what the signal did a long time ago and is going to do in the future should be of no concern; only the present should count.  However to calculate the analytic signal at time $t$ we have to know the signal for all time.}
  
\end{enumerate}

The expression \eqref{eq:f12} obtained for signal \eqref{eq:v12} of the example given in the previous section illustrates well the paradoxes P1 to P4.  It appears that P1 to P4 constitute different hues of the same issue and can be reformulated as follows:

\begin{enumerate}[start=0, label={\textbf{P\arabic*}:}]

\item The range of the spectrum of an analytic signal is, in general, different from the range of the \ac{if} of the same signal.

\end{enumerate}

The fifth paradox, on the other hand, refers to a different but equally crucial inconsistency: the \ac{ht} -- as the \ac{ft} -- requires to calculate the integral of the signal for $t \in \interval$.\footnote{The discrete \ac{ht} transform is obtained, in practice, based on the discrete \ac{ft}.  Hence, the windowing techniques of the short-time discrete \ac{ft} described above to overcome the need for the calculation of an integral for $t \in \interval$ can be applied also to the calculation of the \ac{ht}.  The reader is referred to \cite{Hahn:1996} for more details on the numerical calculation of the \ac{ht}.}  This appears to be a contradiction when one is only interested in the current time $t$.  From a practical point of view, moreover, the knowledge of a signal for $t \in \interval$ can be obviously achieved only through an approximation, e.g., assuming that certain steady-state conditions have been and will be in place forever.

In \cite{Cohen:1995}, there is no attempt to solve these paradoxes, except for an interesting discussion on P5, which argues that the inconsistency arises for the twofold nature -- local and non-local -- of signals \cite{Vakman:1976}.  For example, light is local when interpreted as a particle and non-local when interpreted as a wave.  It is also interesting to note that the paradoxes above are formulated from the point of view of the \ac{ft} or, at least, of the \ac{ht}.  The underlying assumption in the whole \cite{Cohen:1995} as well as in most works on signal processes is that the \ac{ft} is \textit{right}, so one has to reconcile the \ac{if} with it.  However, it would be equally legitimate to discuss the inconsistencies between the \ac{ft} and the \ac{if} of a signal from the point of view of the \ac{if}, i.e.~assuming that the \ac{if} is \textit{right} and trying to reconcile the \ac{ft} with it.

In this work, we do not take the side of either approach.  We propose a geometric framework that assumes that both the \ac{if} (actually, a slightly more general concept, namely the \textit{curvature}, which is introduced in the next section) and the \ac{ft} (or any other time/frequency transform, in fact) are both right and we explain why they seem to provide different information.  This is the topic of the next section.

\section{Geometrical Interpretation}
\label{sec:geometry}

This section approaches the problem of defining the transient behavior of a signal from a completely different point of view with respect to Section \ref{sec:background}.  In the same vein as \cite{Vakman:1976}, one can argue that the underlying approach of the techniques described in Section \ref{sec:background} is to consider the signal a wave, and, as such, to study its properties non-locally, i.e., taking the time as a unique block ranging from $-\infty$ to $\infty$.  One may also argue that the approach described below is intrinsically local, as it studies the properties of signal as a particle the trajectory of which is known at a given position and a given time.  Yet, this interpretation must be reconsidered as the paradoxes described in Section \ref{sub:five1} arise using only non-local approaches such as the \ac{ht}.
The purpose of this section is thus twofold: to introduce first some elementary concepts of differential geometry and then use these concepts to define a common framework where both local and non-local approaches consistently coexist.

\subsection{Space Curves}
\label{sub:curves}

The starting point of any geometry is to define a system of coordinates.  It is easy to accept that the coordinates of a physical space are three.  Extension to a fourth coordinate, time, is less intuitive and a relatively recent extension.  Multi-dimensional spaces defined, for example, in string theories, are even more recent and, for many, quite exotic spaces.

When dealing with electric circuits and power systems, it is much less clear how many dimensions should be considered.  In this work, for simplicity, we consider three, for the three-phase circuits and machines that are typically used in power systems.  It is important to note however that this choice is not a hard limit. Higher dimensions can be taken into account and there exists the math to do that, see e.g. the gentle introduction to frames of arbitrary dimensions given in \cite{Clelland:2017}.

Assuming thus three dimensions as a non-binding constraint, let us consider a space curve $\bfg x(t):[0,+\infty)\rightarrow\mathbb{R}^3$ with $\bfg x(t) = (x_1(t), x_2(t), x_3(t))$.  It is convenient and usual to define an orthonormal basis, say $(\e{1}, \e{2}, \e{3})$, to describe the vector $\bfg x(t)$,
namely:
\begin{equation}
    \bfg x(t) = x_1(t) \, \e{1} + x_2(t) \, \e{2} + x_3(t) \, \e{3} \, .
\end{equation}
A common choice for the basis is the Cartesian coordinates:
\begin{equation}
  \label{eq:cartesian}
  \begin{aligned}
    \e{1} &= (1, 0, 0) \, , \\
    \e{2} &= (0, 1, 0) \, , \\
    \e{3} &= (0, 0, 1) \, . \\
  \end{aligned} \quad \Rightarrow \quad \e{123} = \bfb I_3 \, ,
\end{equation}
but, in turn, any triplet of linearly independent vectors that are orthonormal is perfectly fine.  For example, the rows of the \ac{ct} matrix $\bfb C$ in \eqref{eq:clarke} define a relevant basis in power system analysis.\footnote{Note that $\bfb C$ is not orthogonal.  If a power invariant transform is required, $\bfb C$ is replaced with $\sqrt{\tfrac{3}{2}}\bfb C$, which is orthogonal.}

In differential geometry, it is of particular relevance to define quantities that are \textit{invariant}, that is, do not change when the basis of the coordinates changes.  The most intuitive invariant is arguably the length $s$ of the curve, defined as:
\begin{equation}
  s(t) = \int_0^t \sqrt{\bfp x(r) \cdot \bfp x(r)} \, dr + s_0 \, ,
\end{equation}
from which one obtains the expression:
\begin{equation}
  \label{eq:s}
  s'(t) = \frac{d}{dt} s(t) =
  \sqrt{\bfp x(t) \cdot \bfp x(t)} = |\bfp x(t)| \, ,
\end{equation}
where 
\small
\begin{equation}
  \label{eq:xp}
  \bfg x'(t) = \tfrac{d}{dt} (x_1(t) \, \e{1}(t)) +
  \frac{d}{dt} (x_2(t) \, \e{2}(t)) + \frac{d}{dt} (x_3(t) \, \e{3}(t)) \, ,
\end{equation}
\normalsize
and $\cdot$ represents the inner (or scalar) product of two vectors, hence:
\begin{equation}
  \label{eq:inner}
  \bfg x'(t) \cdot \bfg x'(t) = x^2_1(t) + x^2_2(t) + x_3^2(t) \, .
\end{equation}
Equation \eqref{eq:xp} is written assuming that, in general, the basis $\e{123}(t)$ is time-dependent.  The matrix of the Park transform, say $\bfb P(t)$, is a relevant example of time-dependent basis:
\begin{equation}
  \label{eq:park}
  \bfb P(t) =
  \begin{bmatrix}
    \cos {\left(\vartheta_{\rm p}(t) \right)}&\sin {\left(\vartheta_{\rm p}(t) \right)}&0\\
    -\sin {\left(\vartheta_{\rm p}(t) \right)}&\cos {\left(\vartheta_{\rm p}(t) \right)}&0\\
    0&0&1
  \end{bmatrix}
  \bfb C \, ,
\end{equation}
where $\vartheta_{\rm p}(t) = \omega_{\rm p} t + \theta_{\rm p}$.

According to the chain rule, the derivative of $\bfg x(t)$ with respect to $s(t)$ can be written as:
\begin{equation}
  \label{eq:tangent}
  \bfd x(t) = \frac{d \bfg x (t)}{d s(t)}  =
  \frac{d\bfg x(t)}{dt} \frac{dt}{ds(t)} = \frac{\bfp x(t)}{s'(t)} =
  \frac{\bfp x(t)}{|\bfp x(t)|} \, ,
\end{equation}
where the unit vector $\bfd x(t)$ is tangent to the curve $\bfg x(t)$.

We now define an important moving (e.g., time-dependent) frame, called
Frenet frame, that is defined locally for every point of a smooth curve
$\bfg x(t)$.  This frame is built with the tangent vector, which is
defined in \eqref{eq:tangent}, the normal vector and the binormal
vector, as follows:
\begin{equation}
  \label{eq:TNB}
  \bfb F(t) =
  \begin{bmatrix}
    \T \\ \N \\ \B
  \end{bmatrix} =
  \begin{bmatrix}
    \bfd x(t) \\
    |\bfdd x(t)|^{-1} \, \bfdd x(t)  \\
    \T \times \N   \\
  \end{bmatrix}
  \, ,
\end{equation}
where $\times$ represents the cross product.
%
%
The vectors in \eqref{eq:TNB} are orthonormal, i.e.~$\T = \N \times \B$ and $\N = \B \times \T$, and satisfy the following relations, known as Frenet-Serret formulas \cite{Stoker}:
\begin{equation}
  \label{eq:frenet3}
  \begin{aligned}
    \Td &= \phantom{-}\kappa(t) \, \N \, , \\
    \Nd &= -\kappa(t) \, \T + \tau(t) \, \B \, , \\
    \Bd &= - \tau(t) \, \N \, ,
  \end{aligned}
\end{equation}
where $\kappa(t)$ and $\tau(t)$ are the \textit{curvature} and the
\textit{torsion}, respectively, which are given by:
\begin{equation}
  \label{eq:kappa}
  \kappa(t) = |\bfdd x(t)| = {|\bfp x(t) \times \bfpp x(t)|}\big/{|\bfp x(t)|^3} \, , 
\end{equation}
and
\begin{equation}
  \label{eq:tau}
  \tau(t) = \frac{\bfp x(t) \cdot \bfpp x(t) \times \bfppp x(t)}{|\bfp x(t) \times \bfpp x(t)|^2} \, .
\end{equation}
The quantities defined above, namely $\kappa(t)$ and $\tau(t)$ may vary from point to point but are invariants, like $s(t)$, which means that, while local, do not depend on the coordinates employed to describe the curve.

\subsection{Frequency as an Invariant}
\label{sub:kappa}

As discussed in the introduction, references \cite{freqgeom} and \cite{freqfrenet} present an interpretation of electrical quantities as geometrical invariants of a space curve.  The whole argument is based on two assumptions.  First, the voltage (current) is the velocity of the trajectory of the magnetic flux (electric charge).  This assumption is supported by Faraday's law for the voltage and by the very definition of current intensity as the flow of electric charges through a surface.  The leap of this assumption is that the magnetic flux and electric charge flow can be assumed to be ``curves,'' even though, in practice, they are scalar quantities and, when measured, they can be more naturally thought as signals rather than space curves.  However, if one accepts this assumption, and recalling the definitions given in the previous section, then the following expressions for the voltage can be obtained:
\begin{equation}
  \label{eq:xpv}
  \bfg v(t) \equiv \bfp x(t) \, ,
\end{equation}
and, from \eqref{eq:kappa}:
\begin{equation}
  \label{eq:kappav}
  \kappa(t) = \frac{|\bfg v(t) \times \bfp v(t)|}{|\bfg v(t)|^3} \, ,
\end{equation}
and, from \eqref{eq:tau}:
\begin{equation}
  \label{eq:tauv}
  \tau(t) = \frac{\bfg v(t) \cdot \bfp v(t) \times \bfpp v(t)}{|\bfg v(t) \times \bfp v(t)|^2} \, .
\end{equation}
We note that neither \eqref{eq:kappav} not \eqref{eq:tauv} depend on $\bfg x(t)$, which in this context is the vector of the magnetic flux.  This is good news as the magnetic flux is not an easy quantity to measure or estimate.

Finally, we observe that, from \eqref{eq:tangent} and \eqref{eq:xpv}, one obtains:
\begin{equation}
  \label{eq:sv}
  \frac{d}{dt}s(t) = |\bfg v(t)| \, ,
\end{equation}
which, in a non-relativistic framework, is also an invariant.

Reference \cite{freqfrenet} defines the \textit{azimuthal angular frequency} as:
\begin{equation}
  \label{eq:wk}
  \wk = |\bfg v(t)| \, \kappa(t) \, ,
\end{equation}
which can be interpreted as the angular speed in the plane formed by the vectors $\T$ and $\N$; and the \textit{torsional angular frequency} as:
\begin{equation}
  \label{eq:wt}
  \wt = |\bfg v(t)| \, \tau(t) \, ,
\end{equation}
which can be interpreted as the angular speed in the plane formed by the vectors $\N$ and $\B$.

The torsional angular frequency exists only for curves of dimensions higher than two (i.e., $\tau(t) \equiv 0$ for plane curves).  The examples discussed in \cite{freqfrenet} show that, for three-phase circuits, $\wt \ne 0$ for voltages with unbalanced  phase angles and/or harmonic content.

On the other hand, we observe that, for balanced and positive sequence voltages, the azimuthal angular frequency $\wk$ coincides in effect with the \ac{if} defined in \eqref{eq:if}.  That is, merging \eqref{eq:ct3}, \eqref{eq:if} and \eqref{eq:kappav} and assuming $\bar{v}^-(t) = 0$, then:
\begin{equation}
  \label{eq:point}
  \wk \equiv \vartheta'(t) \equiv \phi'(t) \, .
\end{equation}
Finally, we note that all the formulas given in this section are ``instantaenous'', i.e., utilize only qunatities at a given time $t$.  As opposed to the \ac{ft} and \ac{ht}, thus, these formulas do not require the calculation of integrals for $t \in \interval$.  However, the estimation of $\wk$ and $\wt$ does require the calculation of time derivatives, which poses practical challenges.  This point is further  discussed in Section \ref{sub:ex:39bus}.

\subsection{Revisiting the Paradoxes of the Instantaneous Frequency}
\label{sub:five2}

We are ready to revisit the paradoxes presented in Section \ref{sub:five1}.
The geometric approach discussed above shows that there is, in fact, no inconsistency between \ac{ft} and \ac{if}.  In turn, the two approaches discuss different things.  The frequency of the spectrum is in effect a ``coordinate'' that can be used to represent the signal, whereas \ac{if} is a quantity that represents a property of a curve.  Since there is no point to compare a coordinate with a quantity, paradoxes P0 to P4 are cleared.

The discussion of paradox P5 is more involved.  We start by observing that the \ac{if} can be, under certain conditions, an invariant of the signal for which is calculated.  Assuming that the signal is a curve, the calculation of the curvature depends on the number of coordinates required to represent the curve.  For two-dimensional signals, the curvature (and hence the \ac{if}) is an invariant and represents a local property of the curve. 
For higher-dimensional curves, the curvature is still an invariant but one can also define others.  In turn, a space (or higher-dimension) curve has more than just one frequency.  Space curves have two: $\wk$ and $\wt$.  The azimuthal angular frequency $\wk$ coincides with the common notion of \ac{if} only if the system is balanced and has only the positive sequence.  In this case, in fact, the three-phase signal can be represented as a pair of quantities that is equivalent to an analytic signal.

A single-phase voltage or, more in general, a single signal has an additional issue: the curvature does not exist in one dimension.  Yet, a signal is defined by two (possibly time-variant) quantities: amplitude and phase, which suggests that it can be described in a two-dimensional space.  Thus, the first step is to define a set of coordinates.  This is usually an implicit operation in ac circuit analysis as all phasors are, in effect, analytic signals shifted by the reference angular frequency and referred to a common reference phase angle.  On the other hand, in signal processing, the operation of defining a set of coordinates is provided by the \ac{ht} which generates a coordinate shifted by $-90^{\circ}$ with respect to the coordinate of the signal itself.

This observation allows reconsidering paradox P5.  The fact that one has to calculate \ac{ht} of the signal for $t \in \interval$ has not to be interpreted as an inconsistency of the calculation of the \ac{if}.  The function of the \ac{ht} is just that of allowing the definition of a system of Cartesian coordinates for the signal.  Now, the \ac{ht} dictates that, to be able to define such a coordinate, one needs to know the full signal.  In stationary ac circuits, the knowledge of the full signal is not needed simply because every quantity only has one frequency ($\omega_o$) or multiples of it (harmonics $h \omega_o$).  With this assumption, it is possible to define an absolute reference angle and, hence, a set of coordinates, without the need of calculating the \ac{ht}.

\section{Towards a Unique Geometric Framework}
\label{sec:examples}

The identities given in \eqref{eq:point} are an important result as they constitute the conditions for which the \ac{ht}, analytic signals, \ac{ct} and the geometric approach agree on what is the ``frequency'' of a signal.

To have a common framework that connects all the transforms that we have discussed so far, it remains to accommodate the \ac{ft}.  With this aim, we observe that if $\vartheta(t) = \omega_o t$, then one obtains:
\begin{equation}
  \label{eq:trivial}
  \wk \equiv \vartheta'(t) \equiv \phi'(t) \equiv \omega_o \, ,
\end{equation}
which is an expected result but does not help understand the relationship between the various approaches.  When there is only one constant frequency in a balanced system, then it comes with no surprise that, independently of how one proceeds, the angular frequency is always the same.  One can also argue that the case of unique constant frequency is exactly the case for which one does not need to estimate the frequency at all \cite{8586583}.  And, as a matter of fact, when studying balanced stationary ac circuits with phasors, the angular frequency is not needed in the equations other than for calculating reactances and susceptances.

The remainder of this section discusses a variety of examples, as follows: Section \ref{sub:ex:2omega} revisits the voltage of \eqref{eq:v12} in view of the geometric approach discussed above; Section \ref{sub:ex:transient} discusses the case of a balanced three-phase voltage resembling an electromechanical transient; Section \ref{sub:ex:unbalanced} discusses two relevant cases of stationary unbalanced three-phase voltages; and Section \ref{sub:ex:39bus} compares the frequency estimation of a conventional \ac{pll} with that obtained using the geometric approach.

\subsection{Single Voltage with Two Harmonics}
\label{sub:ex:2omega}

\begin{figure}[!ht]
  \begin{center}
    \subfigure[Space $(\omega, \fourier{v})$]{\resizebox{0.475\linewidth}{!}{\includegraphics{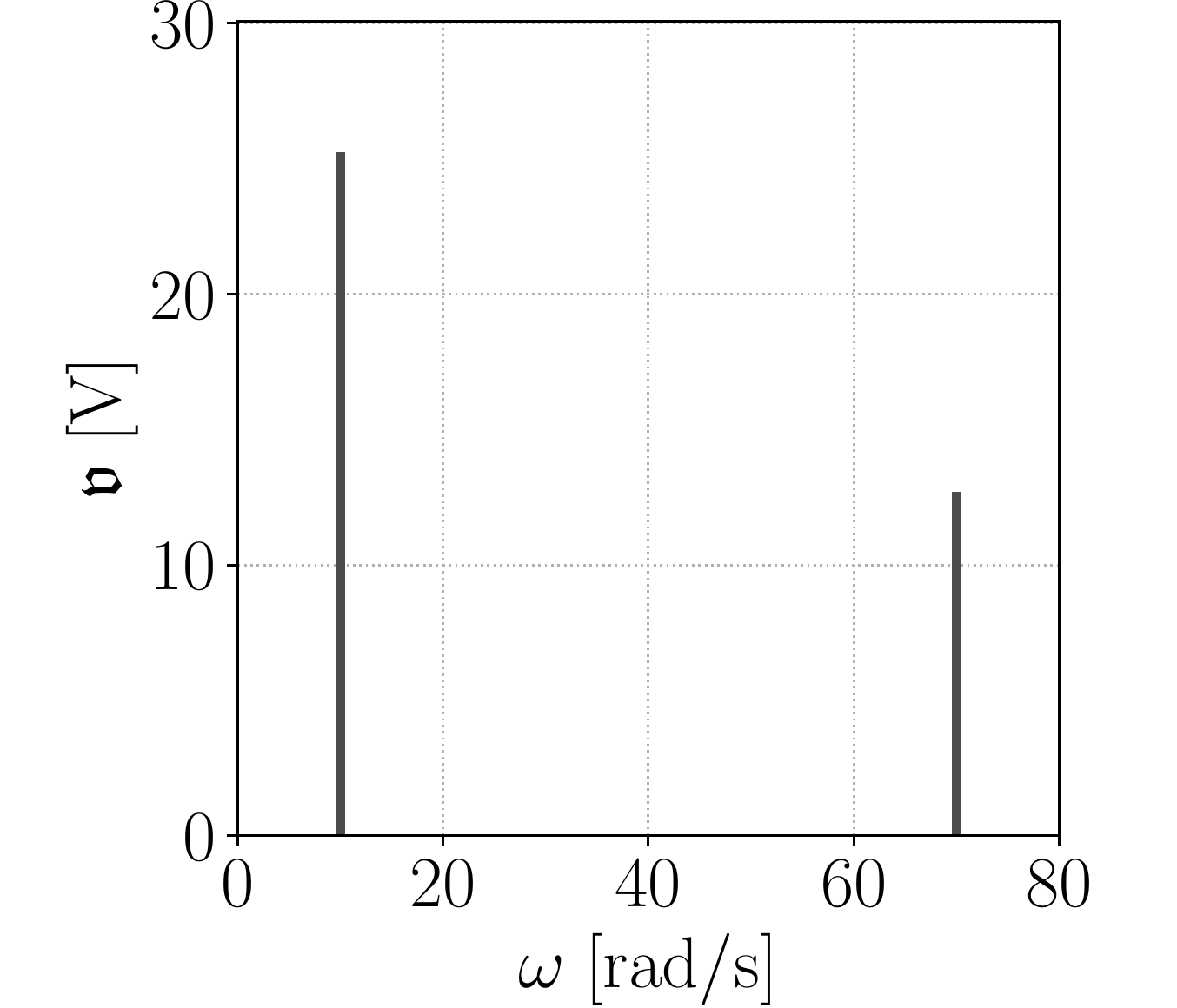}}}
    \subfigure[Space $(v(t), \hilbert{v})$]{\resizebox{0.475\linewidth}{!}{\includegraphics{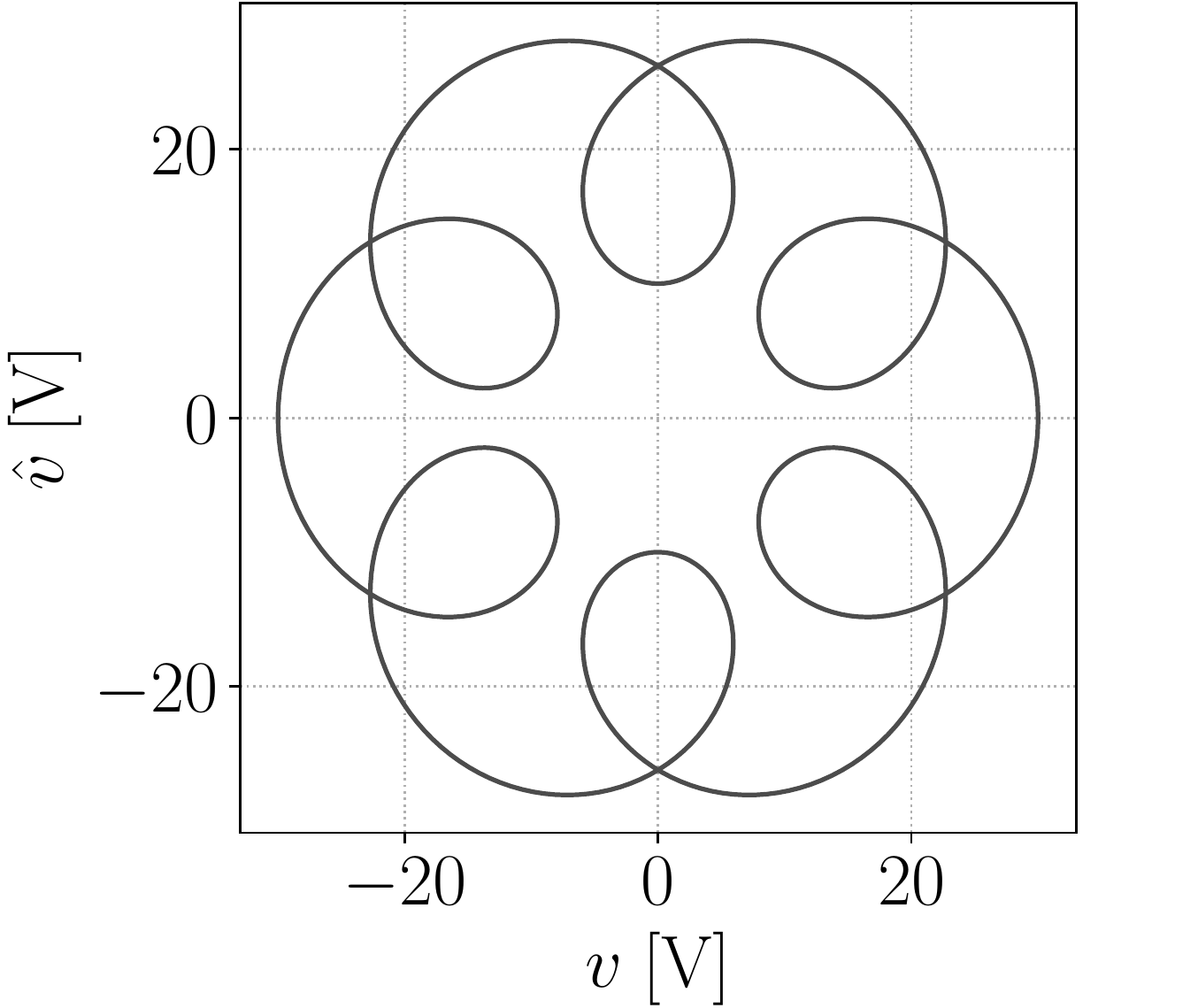}}}
    \caption{Representation of the signal \eqref{eq:v12}.}
    \label{fig:ftht}
  \end{center}
\end{figure}

Let us consider the case for which the voltage contains more than one frequency.  We utilize again the voltage of \eqref{eq:v12}.  As per the \ac{ft}, this signal is represented through a sharp spectrum with two Dirac $\delta$ functions, as shown in Fig.~\ref{fig:ftht}.a.  We can also represent this signal in the state-space $(v(t),\hilbert{v})$, as shown in Fig.~\ref{fig:ftht}.b.  The latter representation is justified from the observation that the \ac{ht} of a signal shifts it by $-90^{\circ}$.

\begin{figure}[!ht]
  \begin{center}
    \resizebox{0.55\linewidth}{!}{\includegraphics{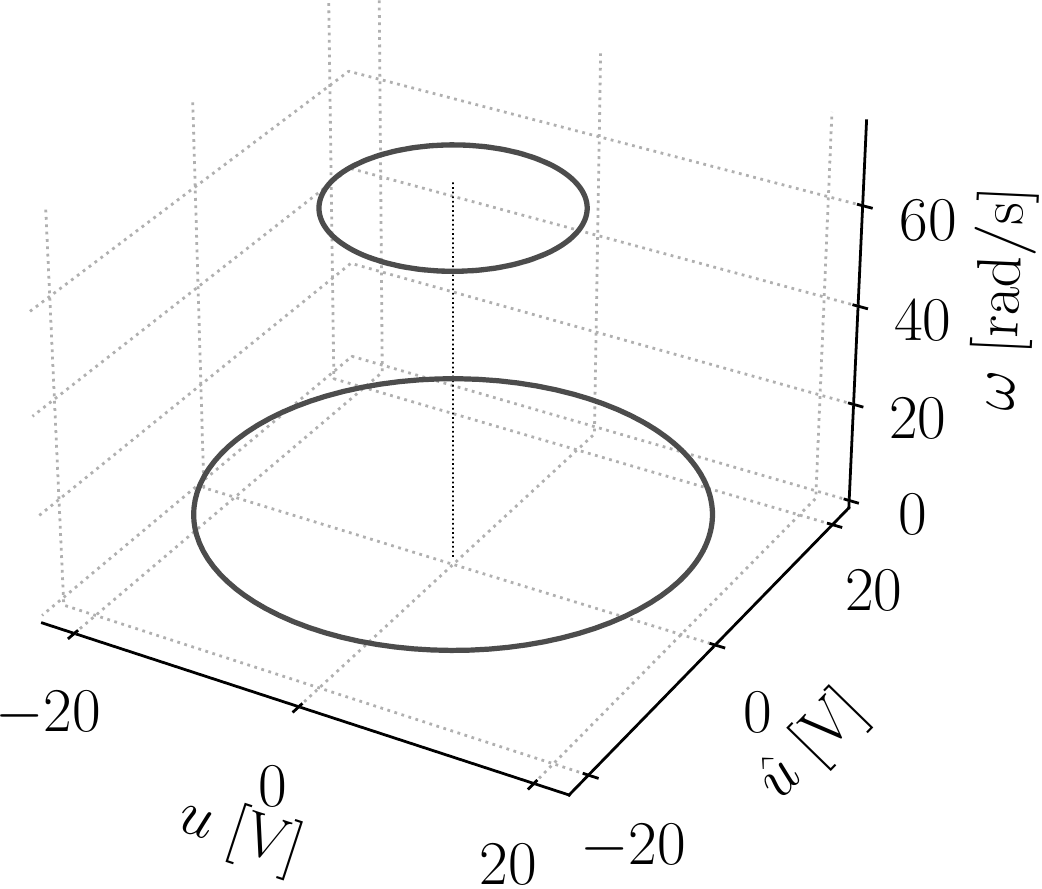}}
    \caption{Representation of the signal \eqref{eq:v12} in the space $(u, \hat{u}, \omega)$.}
    \label{fig:wvh}
  \end{center}
\end{figure}

The proposed framework assumes that the analytic signal and \ac{ft} define a surface mapped by $(t, \omega)$ on the three-dimensional space described by the coordinates $(u, \hat{u}, \omega)$.  The representation of the signal of \eqref{eq:v12} in this space is shown in Fig.~\ref{fig:wvh}.  Since the spectrum of this signal is discrete, there is no actual surface in this case.  For every value of $\omega$, this space represents the content of the analytic signal for $t\in \interval$.  For the signal \eqref{eq:v12}, this content is null except for $\omega = \{ \omega_1, \omega_2 \}$ and the signal can be represented in the space $(u, \hat{u}, \omega)$ as:
\begin{equation}
  \bfg v(t, \omega) =
  \left (
  \begin{matrix}
    v_1(t)\delta(\omega-\omega_1) + v_2(t) \delta(\omega-\omega_2) \\
    \hat{v}_1(t)\delta(\omega-\omega_1) + \hat{v}_2(t) \delta(\omega-\omega_2) \\
    \omega_1 \delta(\omega-\omega_1) + \omega_2 \delta(\omega-\omega_2)
  \end{matrix}
  \right ) ,
\end{equation}
where $v_1(t) = V_1 \cos(\omega_1 t)$ and $v_2(t) = V_2 \cos(\omega_2 t)$.
Most importantly, for every value of $\omega$, the curve is a circle, which has constant curvature and constant \ac{if} equal to the components of $\bfg v(t, \omega)$ on the $\omega$-axis itself.  In this simple example, there are only two circles and the \ac{if} becomes:
\begin{equation}
  \phi'(t, \omega) =
  \omega_1 \, \delta(\omega-\omega_1) +
  \omega_2 \, \delta(\omega-\omega_2) \, .
\end{equation}
On the other hand, the curvature $\kappa(t)$ -- and thus $\wk$, which in this example is given by \eqref{eq:f12} -- of the signal is an invariant and must remain the same, regardless which reference is utilized.  This implies that the orthonormal basis that allows projecting the plane $(v, \hat{v})$ onto the surface defined by the variables $(t, \omega)$ of the space $(u, \hat{u}, \omega)$ is time-dependent and rotates with an angular frequency that is a function of $\wk$.\footnote{The projection of a two-dimensional space onto a surface of a three-dimensional one is a common operation in differential geometry.  This operation is called mapping.  For example, complex numbers can be represented unequivocally on the surface of a sphere through a conformal stereographic projection (see, for example, \cite{Needham:2021}).}  Fortunately, one does not have to find explicitly the equations of this projections, as $\kappa(t)$ and $\wk$ can be obtained directly from \eqref{eq:kappav} and \eqref{eq:wk}, respectively.

Finally, one can view the plots of Fig.~\ref{fig:ftht} with a different perspective.  On the one hand, $\fourier{v}$ is the projection of $\analytic{v}$ onto the $\omega$-axis, which is, in effect, the meaning of the integral \eqref{eq:fourier} that defines the \ac{ft}.  On the other hand, $\analytic{v}$ can be interpreted as the projection of $\fourier{v}$ onto the space $(v, \hat{v})$.

\subsection{Three-phase Voltage Resembling an Electromechanical Transient}
\label{sub:ex:transient}

So far, we have discussed the case of a single signal.  As discussed in Section \ref{sub:clarke}, balanced three-phase systems with only the positive sequence are equivalent to a single signal assuming the use of the space $(u_{\alpha}, u_{\beta}, \omega)$ rather than $(u, \hat{u}, \omega)$.

\begin{figure}[!ht]
  \begin{center}
    \subfigure[Space $(\omega, \fourier{v})$]{\resizebox{0.475\linewidth}{!}{\includegraphics{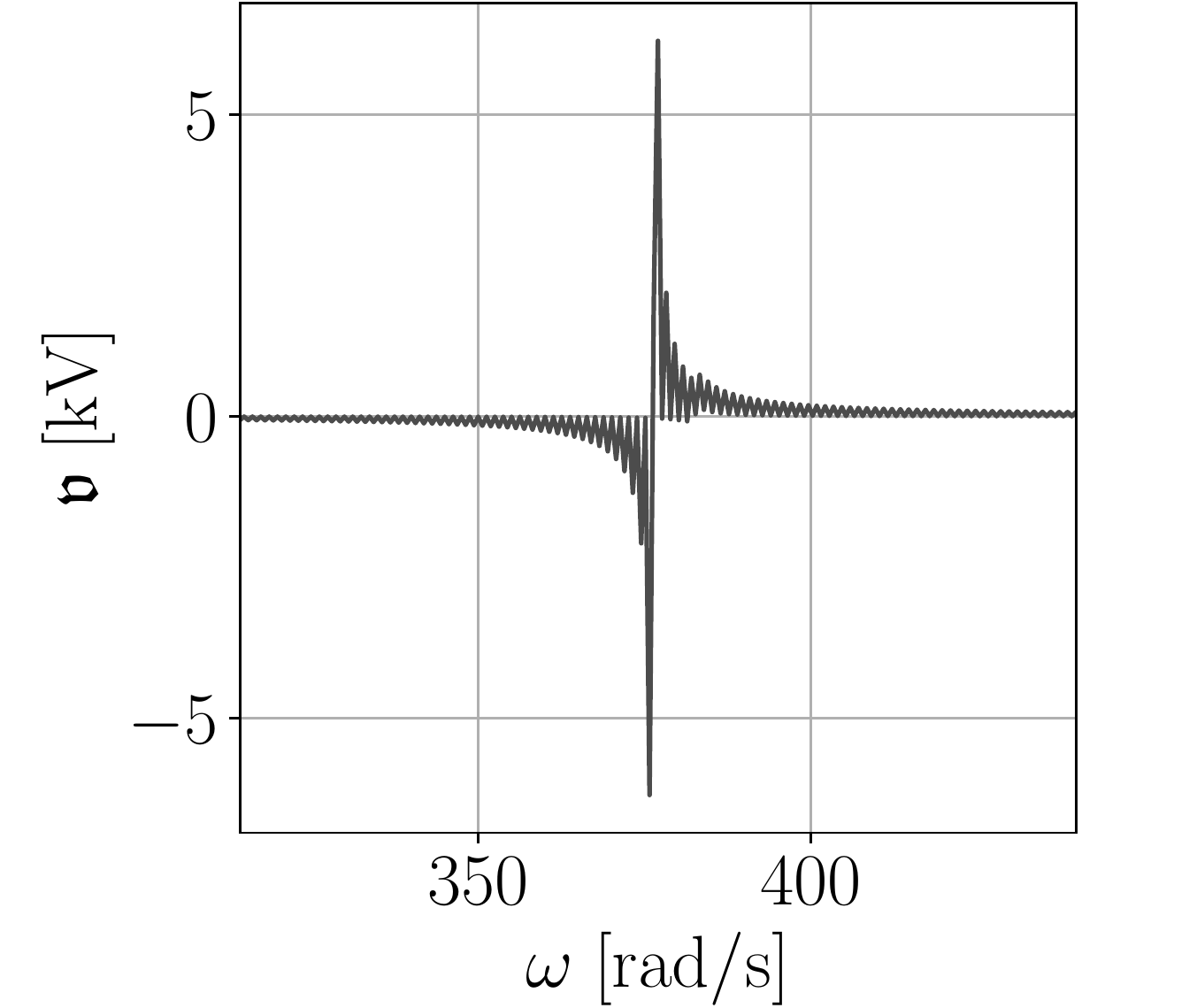}}}
    \subfigure[Space $(v_{\alpha}, v_{\beta})$]{\resizebox{0.475\linewidth}{!}{\includegraphics{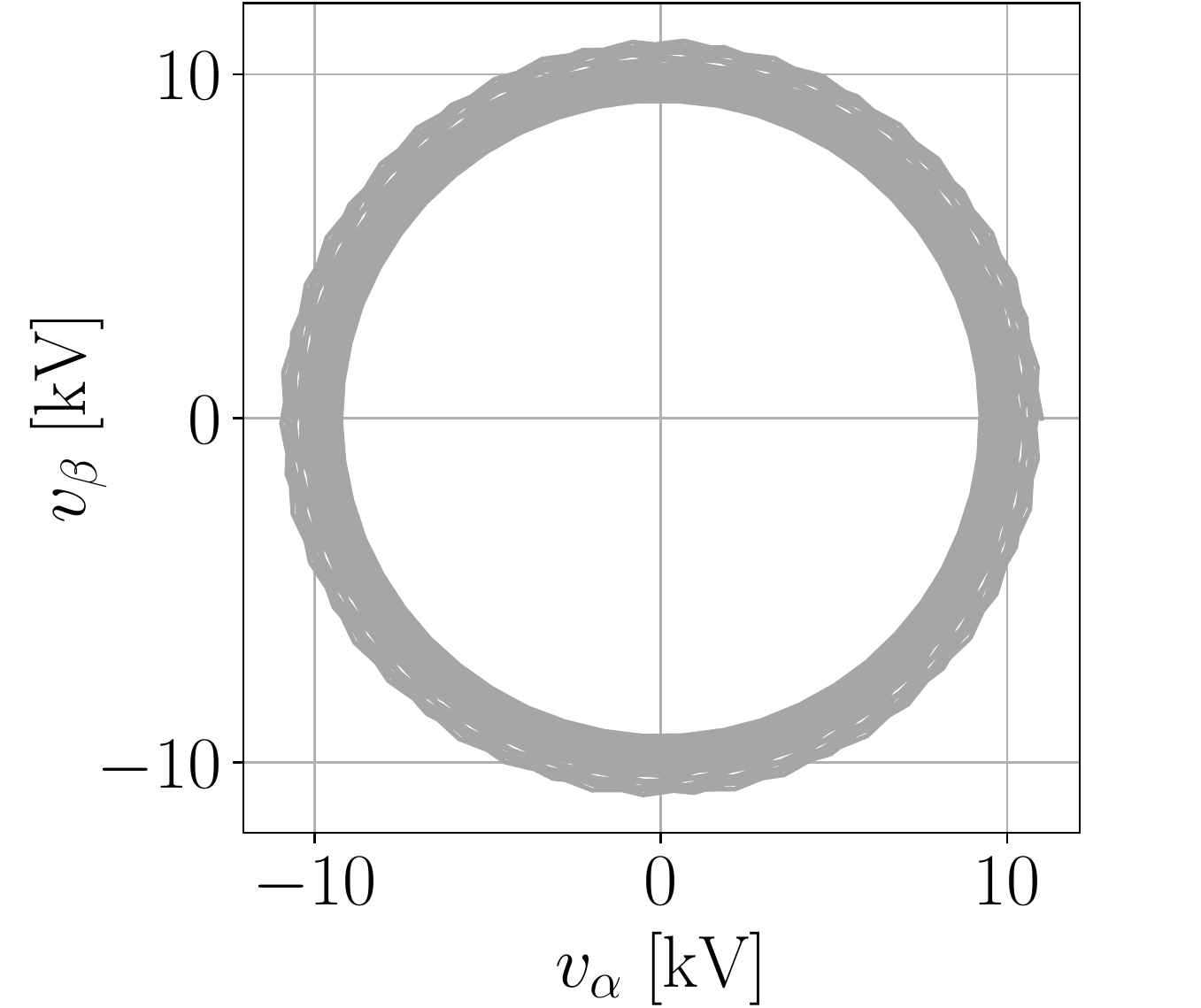}}}
    \caption{Representation of the voltage \eqref{eq:clarke:ex}.}
    \label{fig:park1}
  \end{center}
\end{figure}

To illustrate the proposed framework for a non-stationary case, we consider another example that resembles the dynamic performance of a voltage during a typical electromechanical transient.  To this aim, we consider the following balanced positive-sequence three-phase voltage in \ac{ct} coordinates:
\begin{equation}
  \label{eq:clarke:ex}
  \bfg v_{\alpha \beta \gamma}(t) =
  \left (
    \begin{matrix}
      V_o \, \left (\cos (\omega_o t) + 0.1 \cos(\omega_ot + \psi(t)) \right ) \\ 
      V_o \, \left (\sin (\omega_o t) + 0.1 \sin(\omega_ot + \psi(t)) \right ) \\
      0
    \end{matrix}
  \right ) ,
\end{equation}
where $V_o = 10$ kV and $\omega_o = 2 \pi \, 60$ rad/s and
\begin{equation}
  \psi(t) = e^{-t} \, \cos(0.1 t) \, .
\end{equation}
Figure~\ref{fig:park1} shows the spectrum of the signal obtained with a discrete sine transform in the interval $t \in [0, 5]$ s as well as its representation in the state-space $(v_{\alpha}, v_{\beta})$.  
As expected, the spectrum is concentrated close to $377$~rad/s but does not have a simple representation as the one of the signal \eqref{eq:v12}.

\begin{figure}[!ht]
  \begin{center}
    \subfigure[Space $(v_{\rm d}, v_{\rm q})$]{\resizebox{0.475\linewidth}{!}{\includegraphics{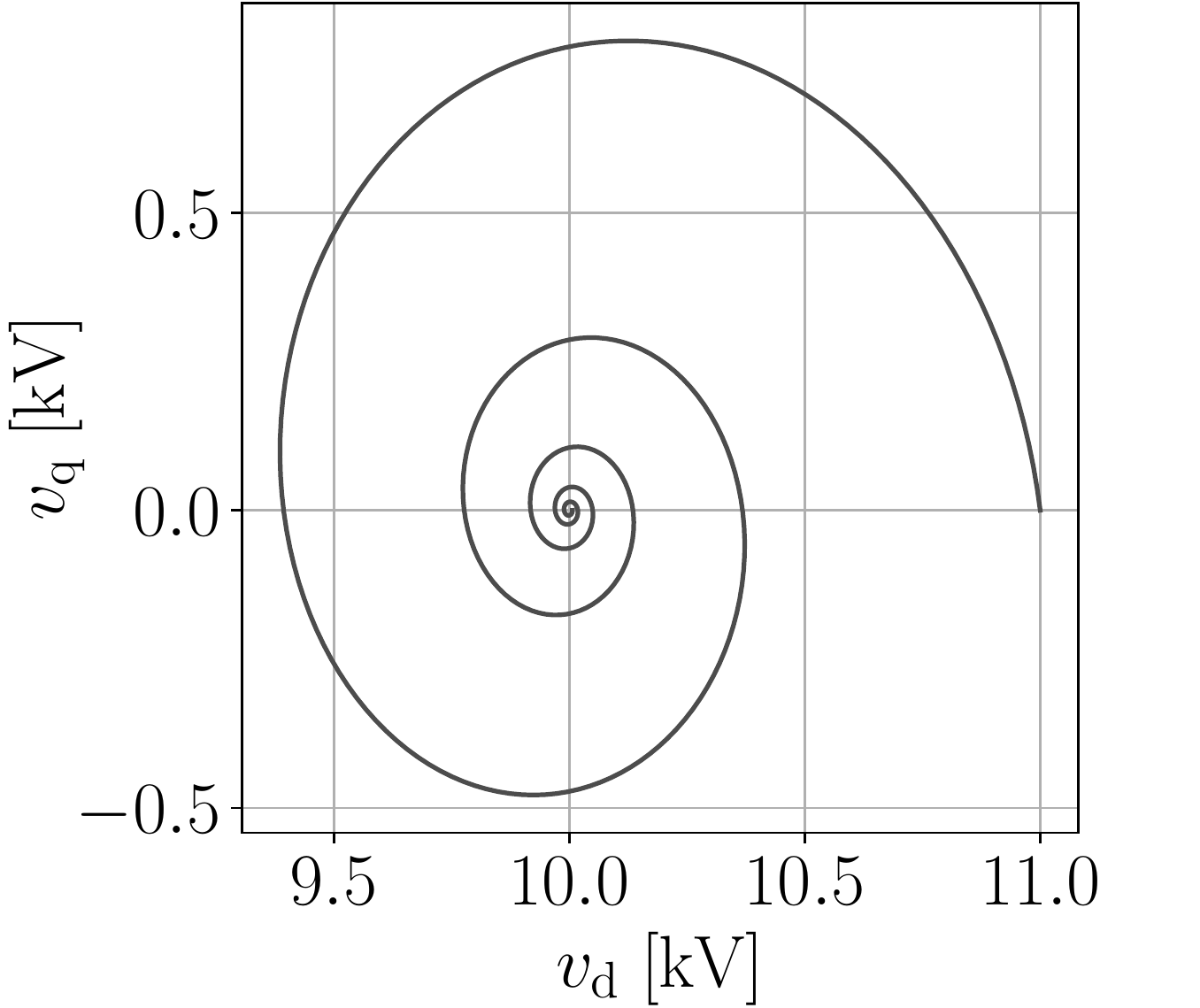}}}
    \subfigure[Instantaneous frequency]{\resizebox{0.475\linewidth}{!}{\includegraphics{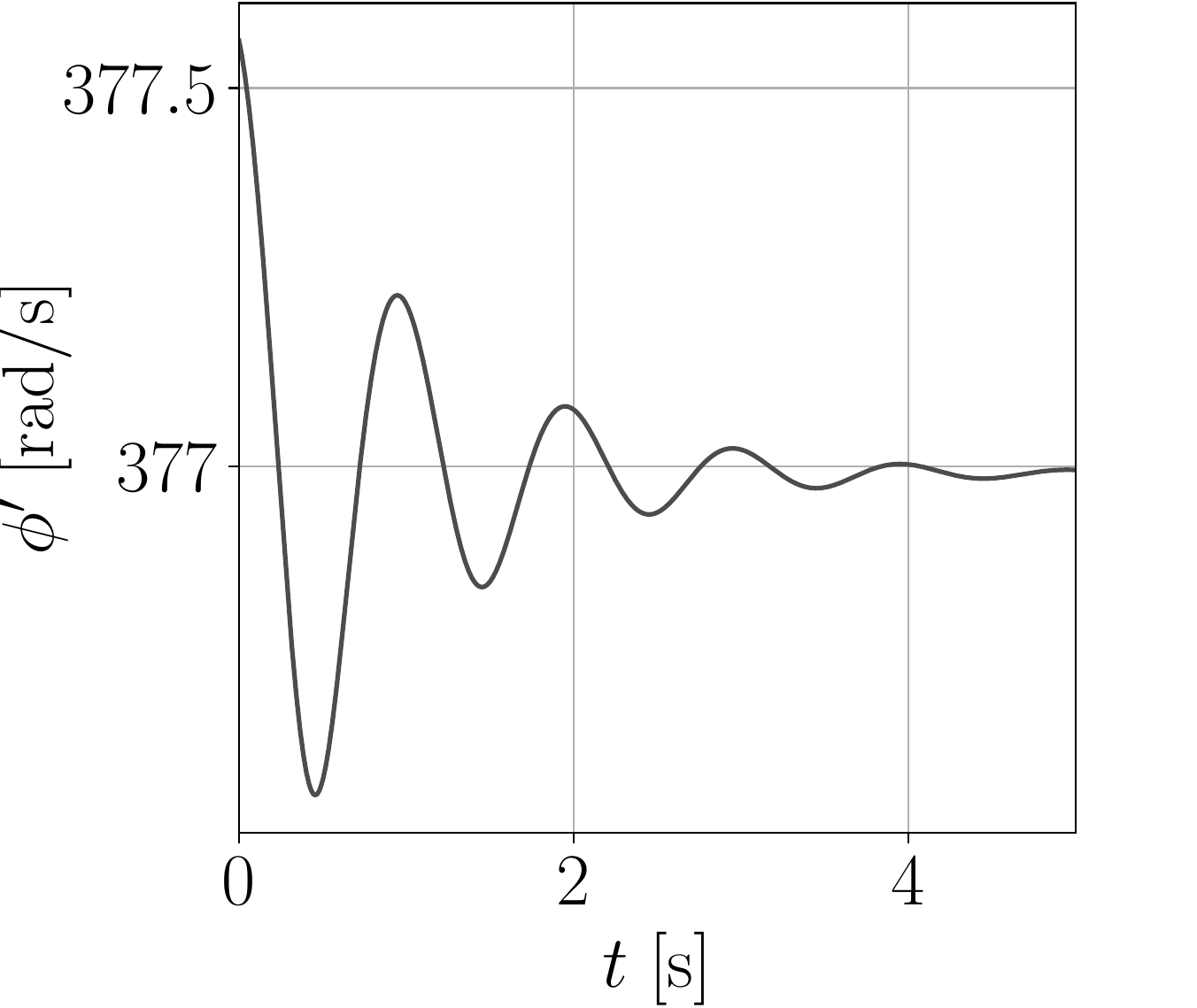}}}
    \caption{Representation of the voltage \eqref{eq:clarke:ex} and its \ac{if}.}
    \label{fig:park2}
  \end{center}
\end{figure}

The \ac{if}, which in this case coincides with $\wk$, can be
calculated from the trajectory in the coordinates
$(v_{\alpha}, v_{\beta})$ with an expression similar to \eqref{eq:if},
as:
\begin{equation}
  \label{eq:clarke:if}
  \phi'(t) = \wk = \frac{v_{\alpha}(t) v'_{\beta}(t) - v'_{\alpha}(t)v_{\beta}(t)}{v^2_{\alpha}(t) + v^2_{\beta}(t)} \, .
\end{equation}
Figure~\ref{fig:park2} shows the \ac{if} obtained with \eqref{eq:clarke:if} as well as the transient behavior of the $\rm dq$-axis component calculated using the Park transform matrix given in \eqref{eq:park} with $\vartheta_{\rm p} = \omega_o t$:
\begin{equation}
  \begin{aligned}
    v_{\rm d}(t) &= 10  + \cos \big ( \psi(t) \big ) \, , \\
    v_{\rm q}(t) &= \sin \big ( \psi(t) \big ) \, .
  \end{aligned}
\end{equation}
Note that in Park coordinates the \ac{if} is the same as  that obtained with Clarke coordinates -- as it has to be -- and has the following expression:
\begin{equation}
  \label{eq:if:park}
  \phi'(t)
  = \omega_o + \frac{v'_{\rm q}(t) v_{\rm d}(t) - v'_{\rm d}(t) v_{\rm q}(t)}{v^2_{\rm d}(t) + v^2_{\rm q}(t)}
  = \omega_o + \phi'_{\rm p}(t) \, .
\end{equation}
The curve shown in Fig.~\ref{fig:park2}.a has \ac{if} equal to $\phi'_{\rm p}(t)$ but its $\wk$ is invariant as the $\rm dq$-axes are rotating with angular speed $\omega_o$.

\begin{figure}[!ht]
  \begin{center}
    \resizebox{0.55\linewidth}{!}{\includegraphics{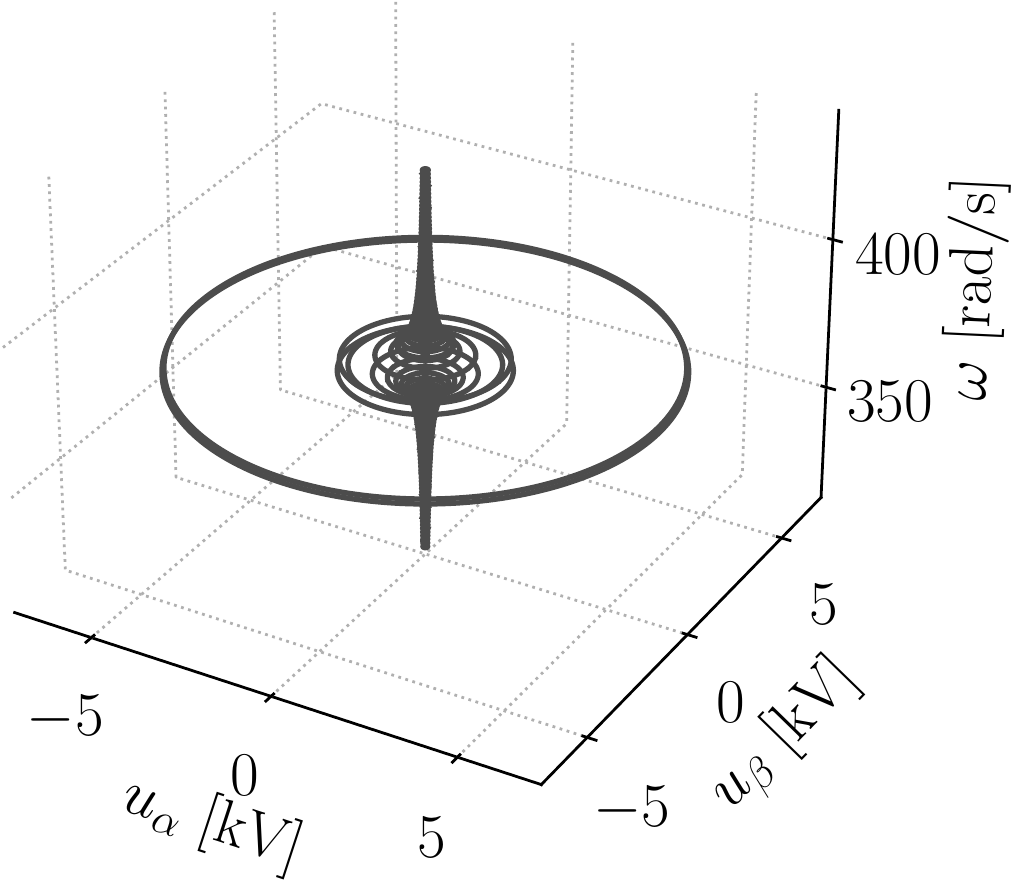}}
    \caption{Representation of the voltage \eqref{eq:clarke:ex} in the
      space $(u_{\alpha}, u_{\beta}, \omega)$.}
    \label{fig:park3}
  \end{center}
\end{figure}

In the same vein as the previous example, one can represent the voltage in a three-dimensional space $(u_{\alpha}, u_{\beta}, \omega)$.  This is shown in Fig.~\ref{fig:park3}.  The information given in these coordinates is the same as that in Figs.~\ref{fig:park1} and \ref{fig:park2}.  However, since the harmonic content of the voltage is not trivial, the information that can be obtained from this representation 
is not straightforward.  The best representation is, in this case, that provided by the Park transform, which explains its common utilization in the study of electromechanical transients of power systems.

It appears, thus, that the spectrum of a transient voltage is not particularly representative of the transient itself.  There exist, of course, a variety of techniques to properly window the measurements of the voltage to obtain an estimation of its \ac{if} (see, e.g., \cite{8675542}) but all of these techniques have the intrinsic limitation that the \ac{ft} is not suited for non-stationary signals.  

Finally, we note that the $\omega$ coordinate introduced by the \ac{ft} is only incidentally a frequency.  Other transforms use functions other than sines and cosines, and may be characterized by different parameters.   For example, wavelets are based on the ${\rm sinc}$ function and the wavelet decomposition consists in finding a scale $a$ and a shift factor $b$ of each wavelet that form the original signal \cite{chui2016introduction}.

\subsection{Unbalanced Three-phase Voltage}
\label{sub:ex:unbalanced}

\begin{figure}[!ht]
  \begin{center}
    \subfigure[Clarke components]{\resizebox{0.475\linewidth}{!}{\includegraphics{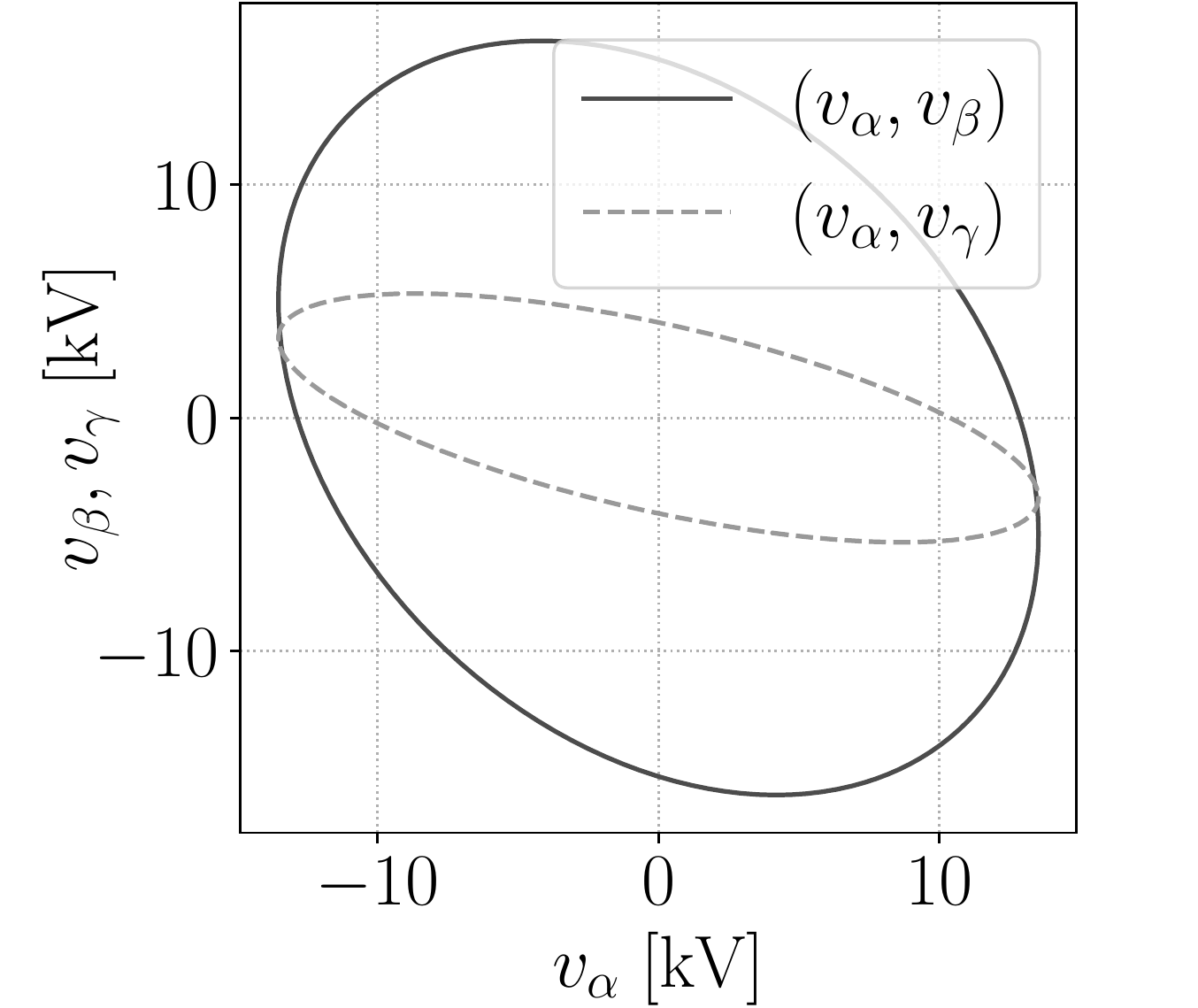}}}
    \subfigure[Angular frequency]{\resizebox{0.475\linewidth}{!}{\includegraphics{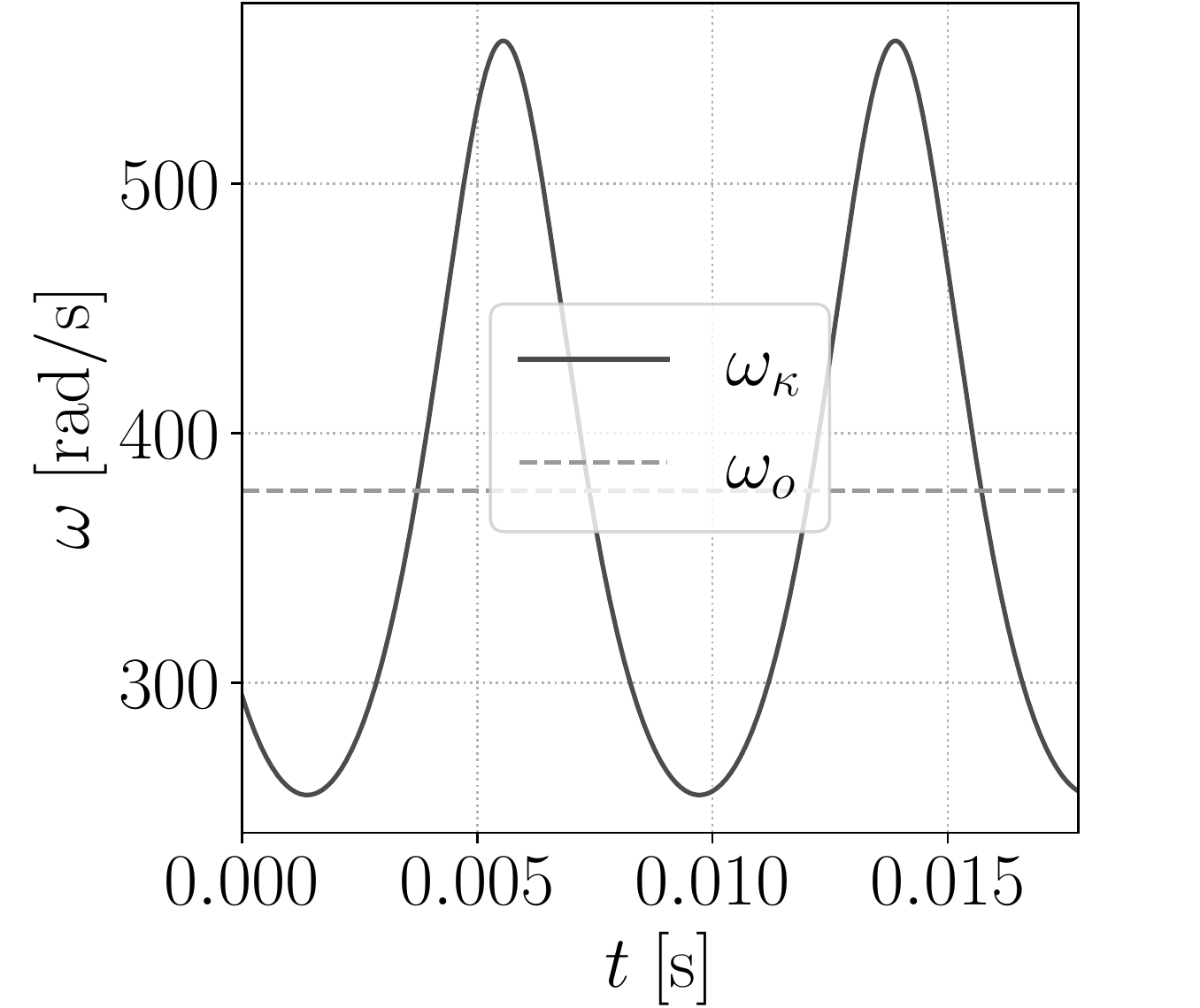}}}
    \caption{Representation of the voltage \eqref{eq:ex:unbalanced} and its $\wk$ and \ac{if}.}
    \label{fig:clarke2}
  \end{center}
\end{figure}

The examples so far have shown cases for which $\phi'(t) = \wk$.  This is, however, not always the case.  The simplest scenario for which the azimuthal frequency is not as expected is a voltage vector with unbalanced amplitudes.  Let us consider the following example:

\begin{equation}
  \label{eq:ex:unbalanced}
  \begin{aligned}
    v_a(t) &= 12 \cos (\omega_o t) \, , \\
    v_b(t) &= 20 \cos (\omega_o t - \tfrac{2}{3}\pi) \, , \\
    v_c(t) &= 12 \cos (\omega_o t + \tfrac{2}{3}\pi) \, ,
  \end{aligned}
\end{equation}
where the amplitudes are in kV and $\omega_o = 2\pi\, 60$ rad/s.  One can readily observe that $\wt = 0$, hence the curve described by $\bfg v_{abc}(t)$ lies in a plane.  However, this plane is not $(\alpha, \beta)$ as $v_{\gamma}(t) \ne 0$ (see Fig.~\ref{fig:clarke2}.a).  Moreover, the curve is not a circle, because $\wk$ is not constant (see Fig.~\ref{fig:clarke2}).  The curve is in fact an ellipse with periodic $\wk$.  This result is consistent with the definition of $\wk$ but, of course, it is not the expected result.  The issue is that $\bfg v_{abc}(t)$ contains positive, negative and zero sequences.  Applying the Fortescue symmetrical component transform in phasor domain, in fact, one obtains:
\begin{equation}
  \begin{aligned}
    \bar{v}^o &= -4 - \jj \, 6.93 \, , \\
    \bar{v}^+ &= -4 + \jj \, 6.93 \, , \\
    \bar{v}^- &= 44 + \jj \, 0 \, ,
  \end{aligned}
\end{equation}
and, thus, $\bfg v_{abc}(t)$ is not an analytic signal.  Only after filtering the negative and zero sequences, one can obtain the expected result, namely $\phi'(t) = \omega_o$.  Finally, note that, in this case, the spectrum coincides with the expected value of the \ac{if}.

Let us now show how unbalanced phases lead to a nonnull torsional frequency.  Consider the three-phase voltage:
\begin{equation}
  \label{eq:ex:torsion}
  \begin{aligned}
    v_a(t) &= 12 \cos (\omega_o t + 0.05 \pi \sin(0.6 t) ) \, , \\
    v_b(t) &= 12 \cos (\omega_o t + 0.05 \pi \sin(0.6 t) - \tfrac{2}{3}\pi) \, , \\
    v_c(t) &= 12 \cos (\omega_o t + d \, 0.05 \pi \sin(0.6 t) + \tfrac{2}{3}\pi) \, .
  \end{aligned}
\end{equation}
\begin{figure}[!b]
  \begin{center}
    \subfigure[Azimuthal frequency]{\resizebox{0.475\linewidth}{!}{\includegraphics{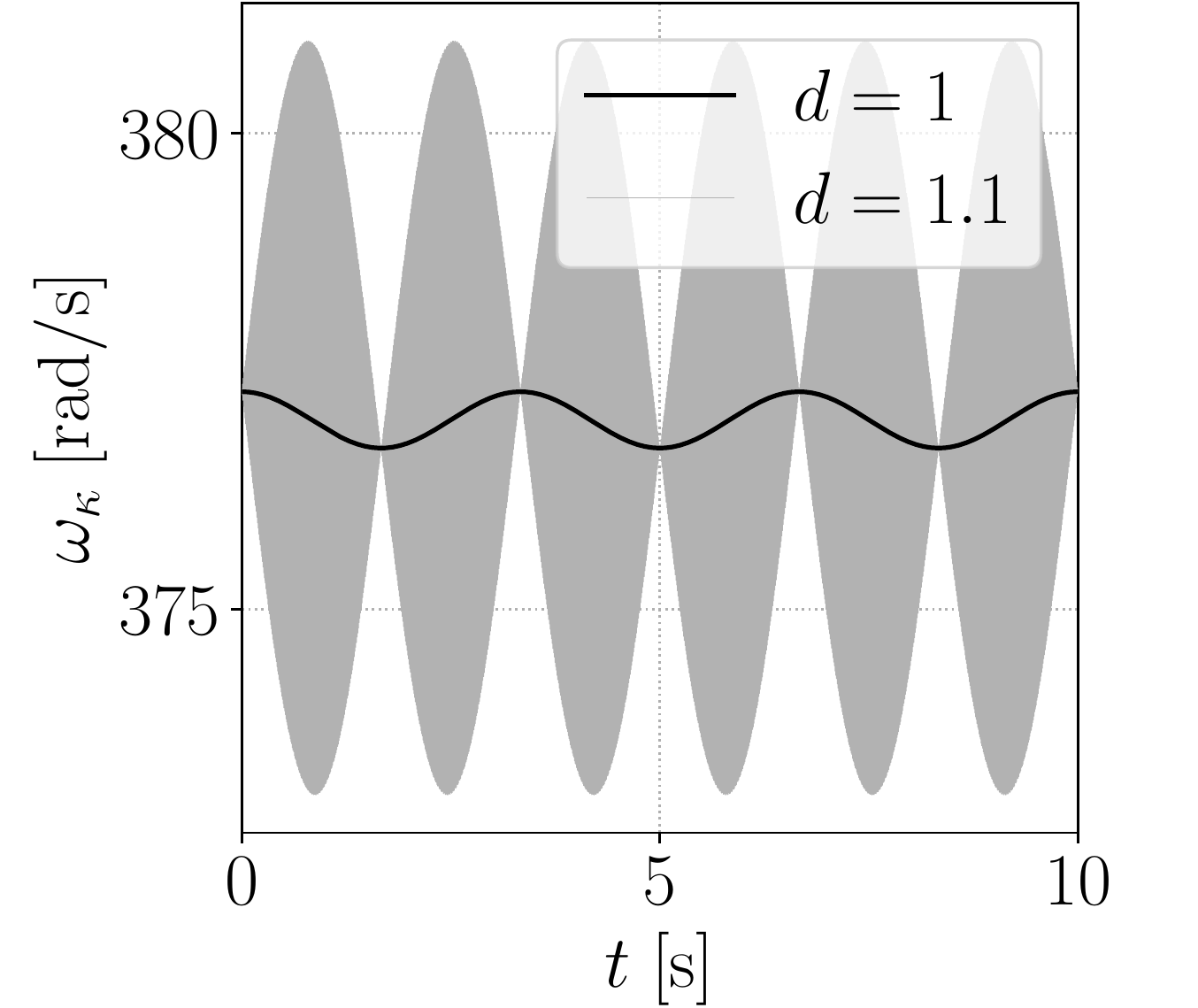}}}
    \subfigure[Torsional frequency]{\resizebox{0.475\linewidth}{!}{\includegraphics{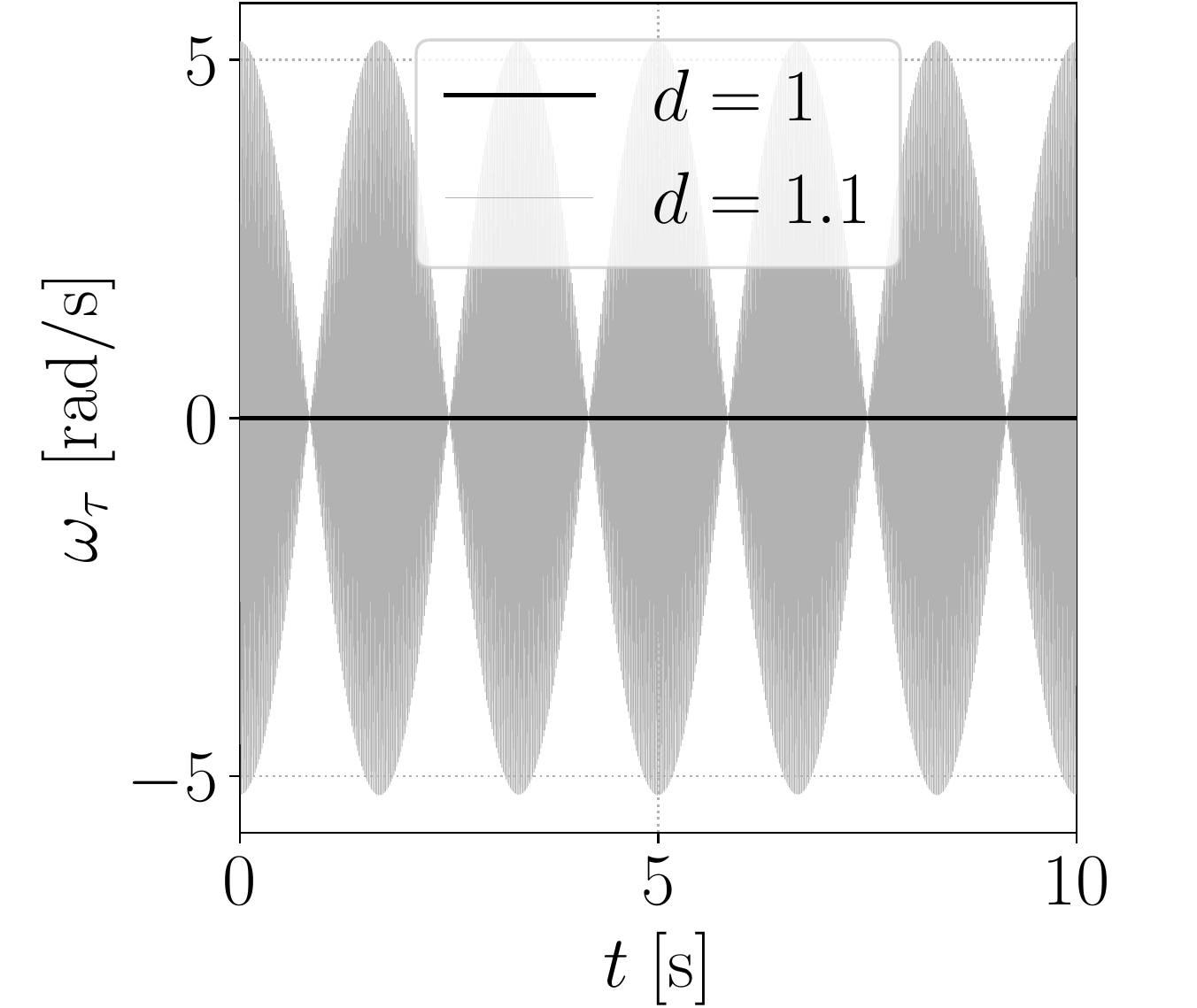}}}
    \caption{Representation of $\wk$ and $\wt$ of the voltage \eqref{eq:ex:torsion}.}
    \label{fig:torsion}
  \end{center}
\end{figure}
Figure \ref{fig:torsion} shows two cases: $d = 1$ and $d=1.1$.  The latter introduces an imbalance in the voltages that leads to high-frequency oscillations of $\wk$ and $\wt$.  These oscillations can be easily filtered.  However, the two examples discussed in this section pose the question whether, in real-world applications, small imbalances can significantly distort $\wk$.  This point is further discussed in the following section. 

\subsection{Frequency during Power System Transients}
\label{sub:ex:39bus}

In high-voltage transmission systems, harmonics and imbalances are minimized by design and proper filtering, in order to comply with network codes.  One has thus to expect that, in most cases, $\wk$ and the \ac{if} return fairly similar values.
In this section, we support this statement by comparing the estimation of the frequency as obtained with a \ac{pll} and the one obtained with \eqref{eq:kappav} starting from the voltage vector $\bfg v_{abc}(t)$ at the bus of a high-voltage transmission system.  

\begin{figure}[!h]
  \begin{center}
    \resizebox{0.7\linewidth}{!}{\includegraphics{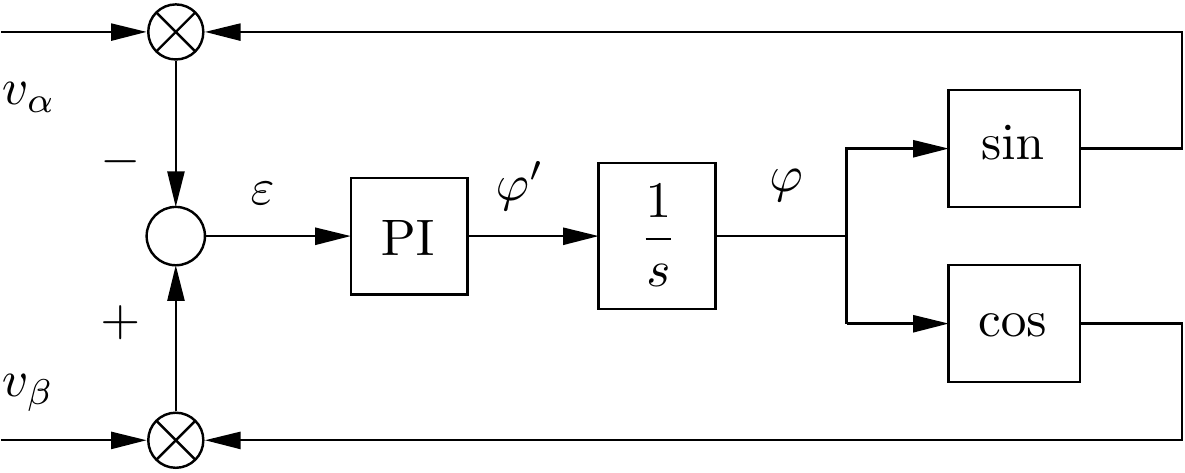}}
    \caption{Scheme of the synchronous reference frame \ac{pll}.}
    \label{fig:pll}
  \end{center}
\end{figure}

The \ac{pll} estimates the \ac{if} based on \eqref{eq:ct2}, as follows.  Assuming $v_{\alpha}(t)$ and $v_{\beta}(t)$ at the bus of interest are known (or measured) and according to the expresison of phase angle of an analytic signal given in \eqref{eq:as4}, one has:
\begin{equation}
  \label{eq:ct:phi}
  \frac{v_{\beta}(t)}{v_{\alpha}(t)} = \frac{\sin(\phi(t))}{\cos(\phi(t))} \, ,
\end{equation}
or, equivalently:
\begin{equation}
  \label{eq:ct:phi2}
  v_{\beta}(t) \cos(\phi(t)) - v_{\alpha}(t) \sin(\phi(t)) = 0 \, .
\end{equation}
The \ac{pll} does not calculate $\phi(t)$ from \eqref{eq:ct:phi2} but estimates it, say $\varphi(t)$, and then tracks the error:
\begin{equation}
  \label{eq:pll:error}
  \varepsilon(t) = v_{\beta}(t) \cos(\varphi(t)) - v_{\alpha}(t) \sin(\varphi(t)) \, .
\end{equation}
Among the many implementations of \acp{pll}, the one utilized in this case study is the synchronous reference frame model, which is shown in Fig.~\ref{fig:pll}.  The input to the integrator, $\varphi'(t)$, is the sought estimation of the \ac{if} \cite{Teodorescu:2011}.  In the simulations, the proportional and integral gains of the PI control of the \ac{pll} are set to $10$ and $30$, respectively.

To carry out the comparison, we consider the fully-fledged EMT model of the IEEE 39-bus system provided by DIgSILENT PowerFactory.  The system model is based on the original IEEE 39-bus benchmark network and is modified to capture the behavior during electromagnetic transients of the power network, namely, the frequency dependency of transmission lines and the non-linear saturation of transformers. For reference this model is available as an application example with DIgSILENT PowerFactory.

\begin{figure}[!h]
  \begin{center}
    \resizebox{0.7\linewidth}{!}{\includegraphics{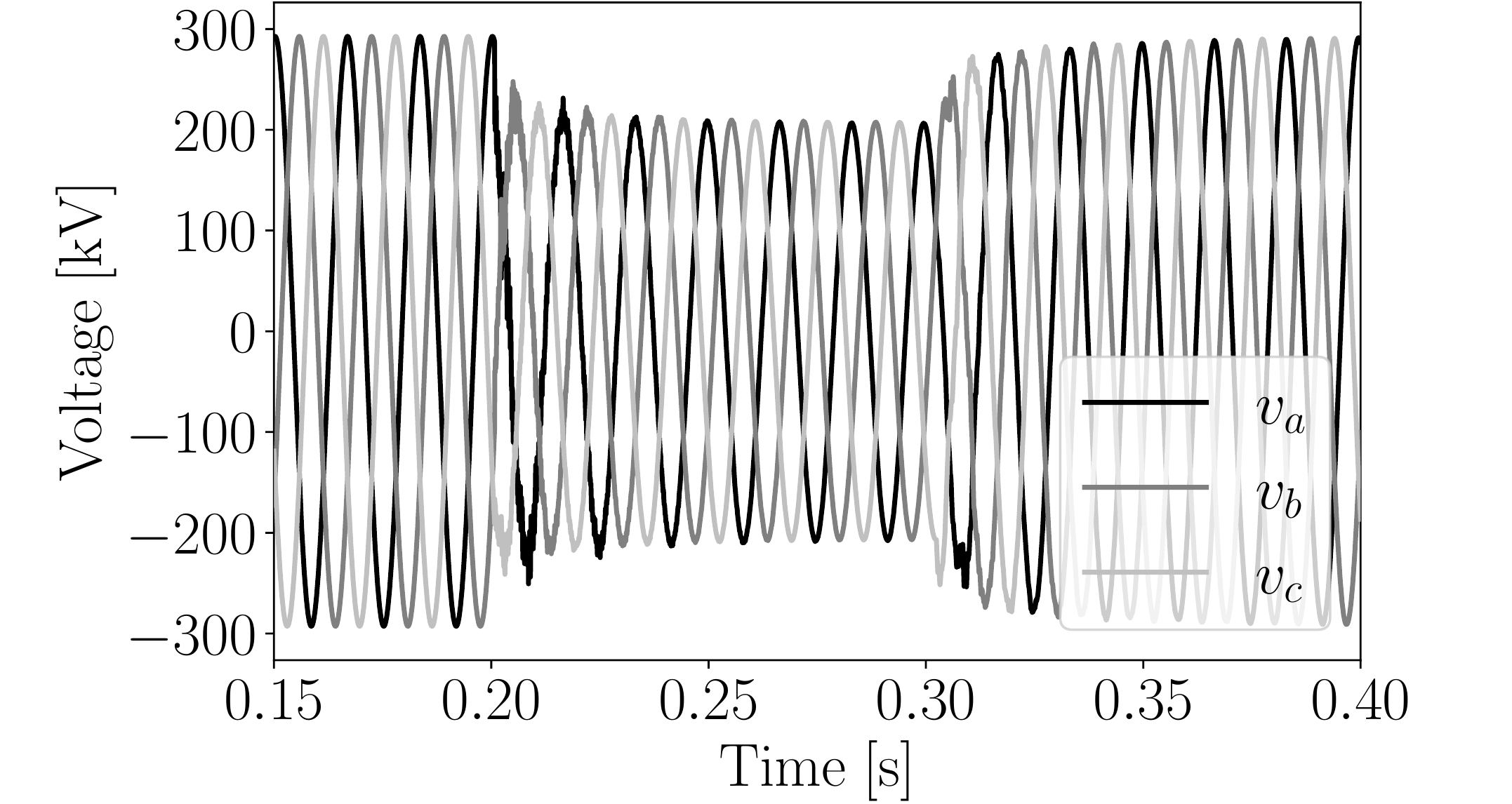}}
    \caption{IEEE 39-bus system, balanced scenario: voltage at bus~26 following the fault at bus 4.}
    \label{fig:volt}
  \end{center}
\end{figure}

A three-phase fault is simulated at terminal bus~4 of the system at $0.2$~s and cleared at $0.3$~s.  The integration time step considered is $0.01$~ms.  We consider the voltages at bus~26 following the contingency.  Figure \ref{fig:volt} shows the trajectories in time of three-phase voltages.

\begin{figure}[!ht]
  \begin{center}
    \subfigure[]{\resizebox{0.475\linewidth}{!}{\includegraphics{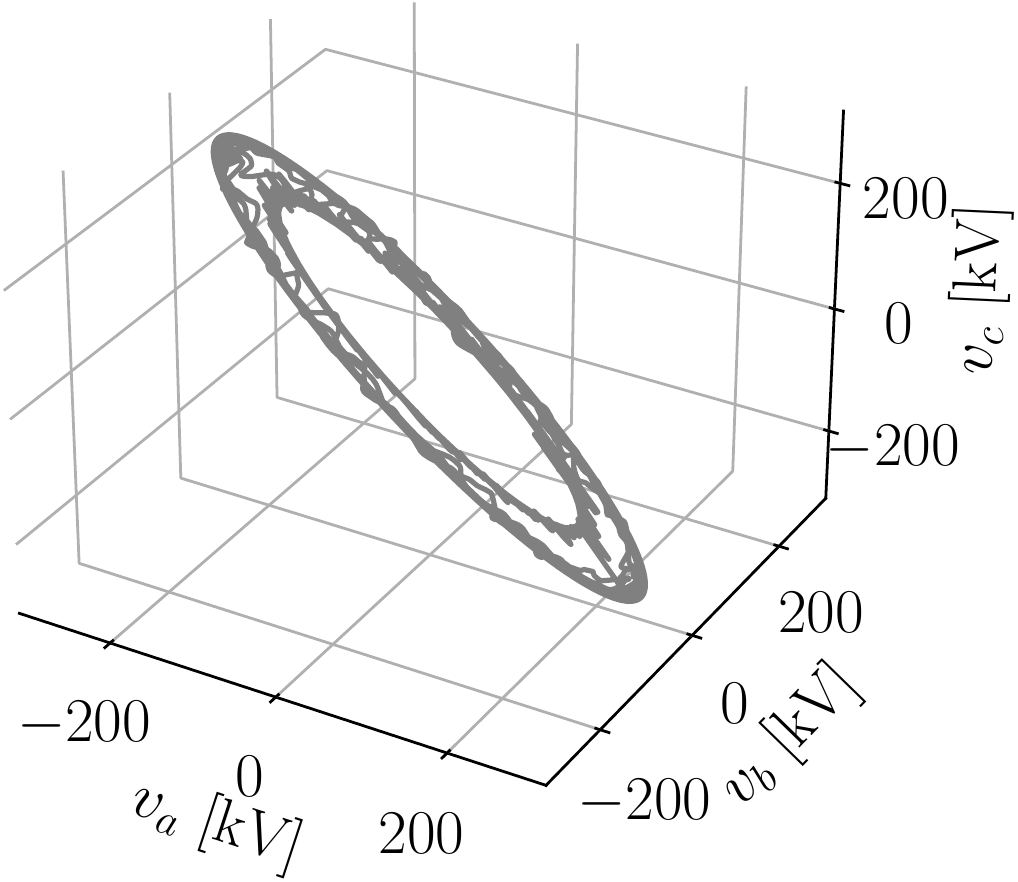}}}
    \subfigure[]{\resizebox{0.475\linewidth}{!}{\includegraphics{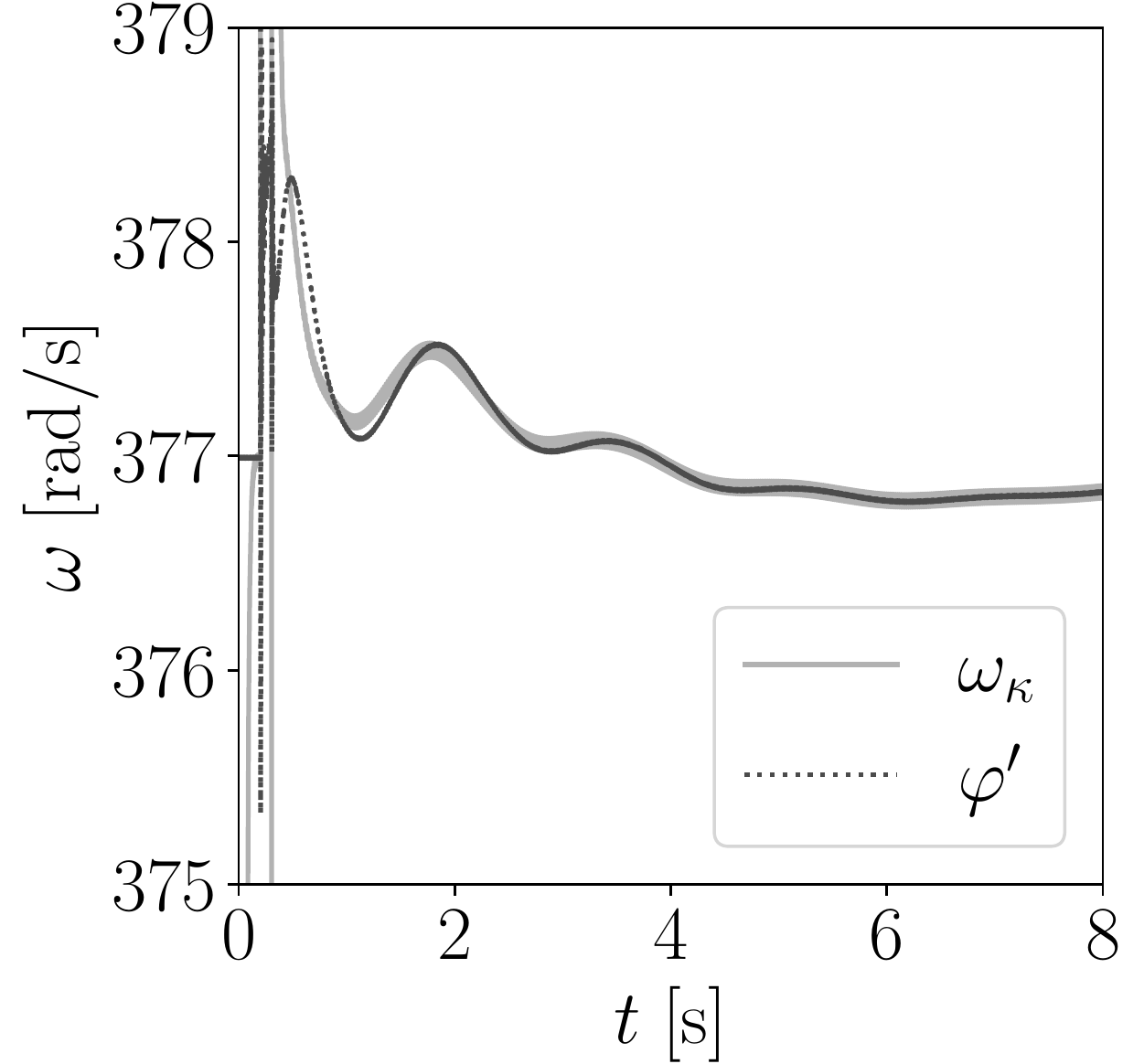}}}
    \subfigure[]{\resizebox{0.475\linewidth}{!}{\includegraphics{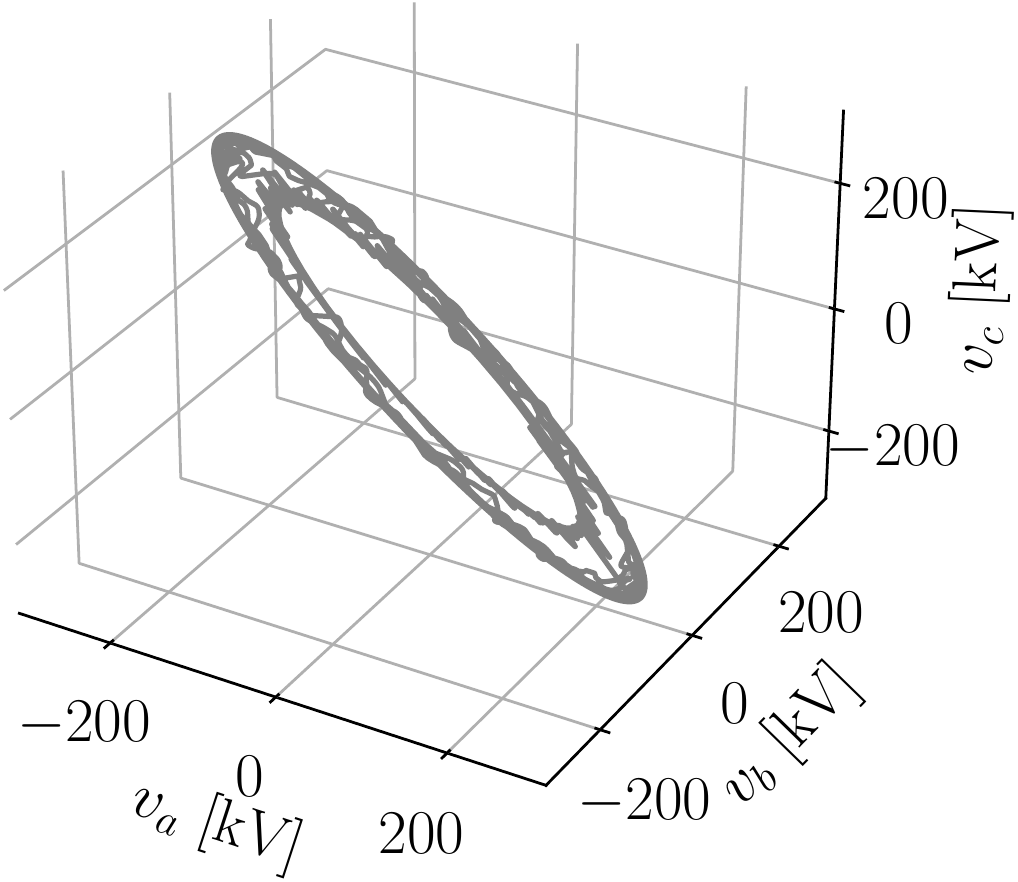}}}
    \subfigure[]{\resizebox{0.475\linewidth}{!}{\includegraphics{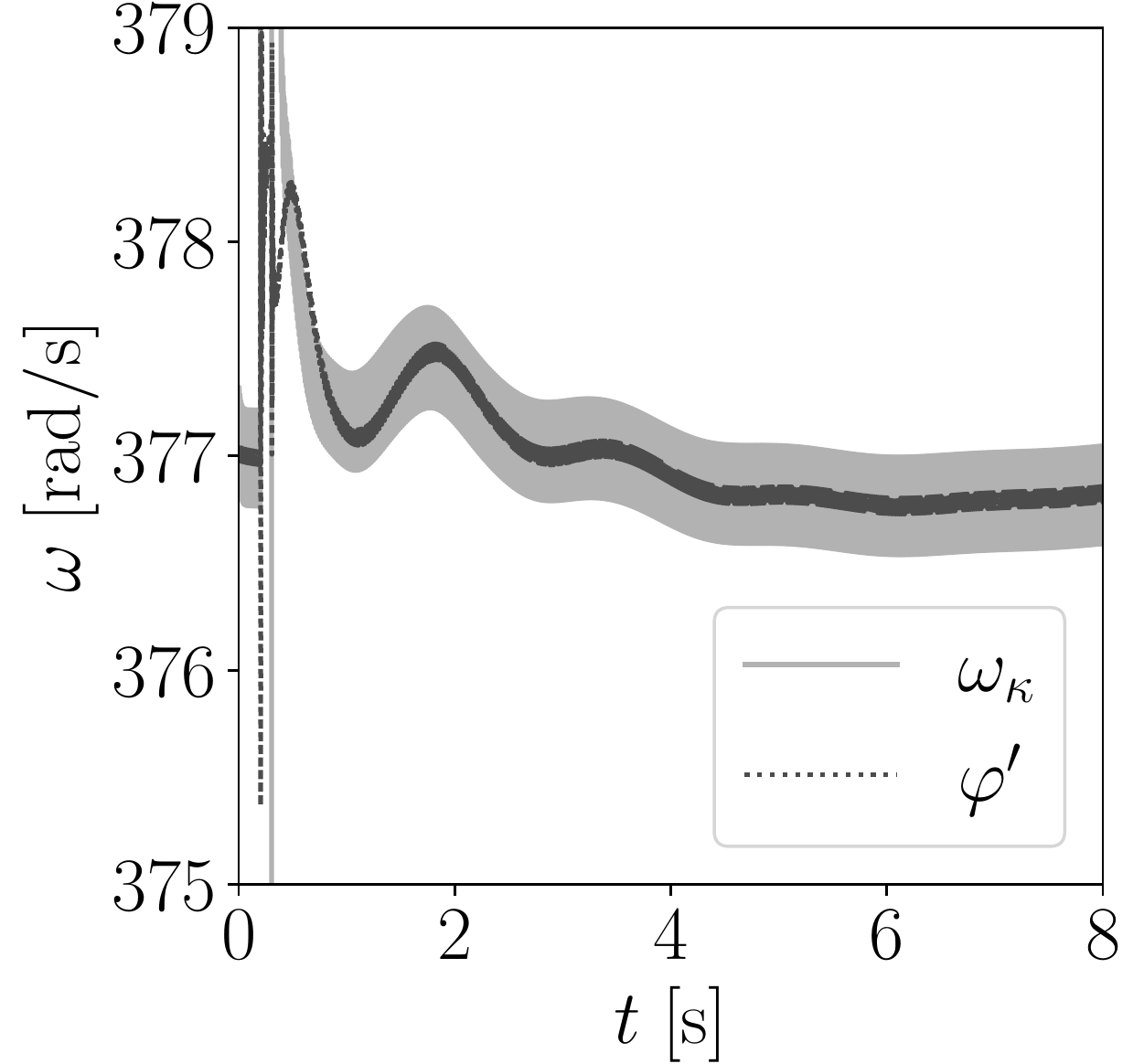}}}
    \subfigure[]{\resizebox{0.475\linewidth}{!}{\includegraphics{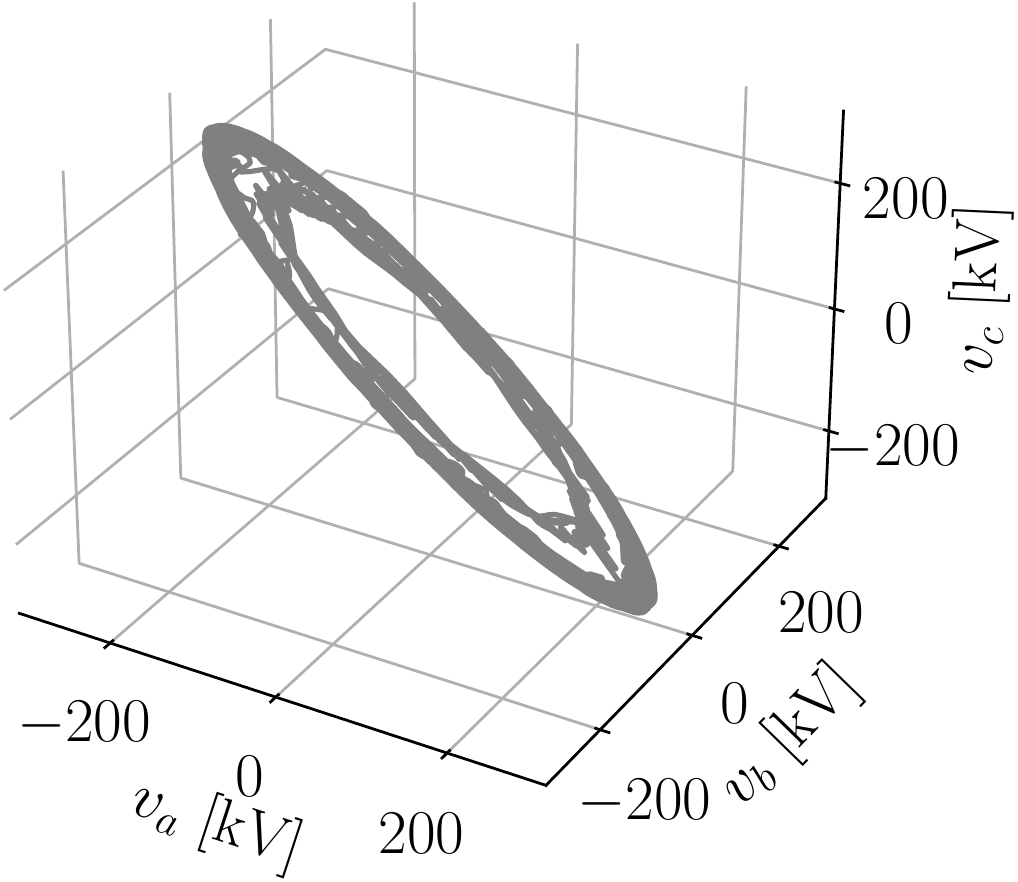}}}
    \subfigure[]{\resizebox{0.475\linewidth}{!}{\includegraphics{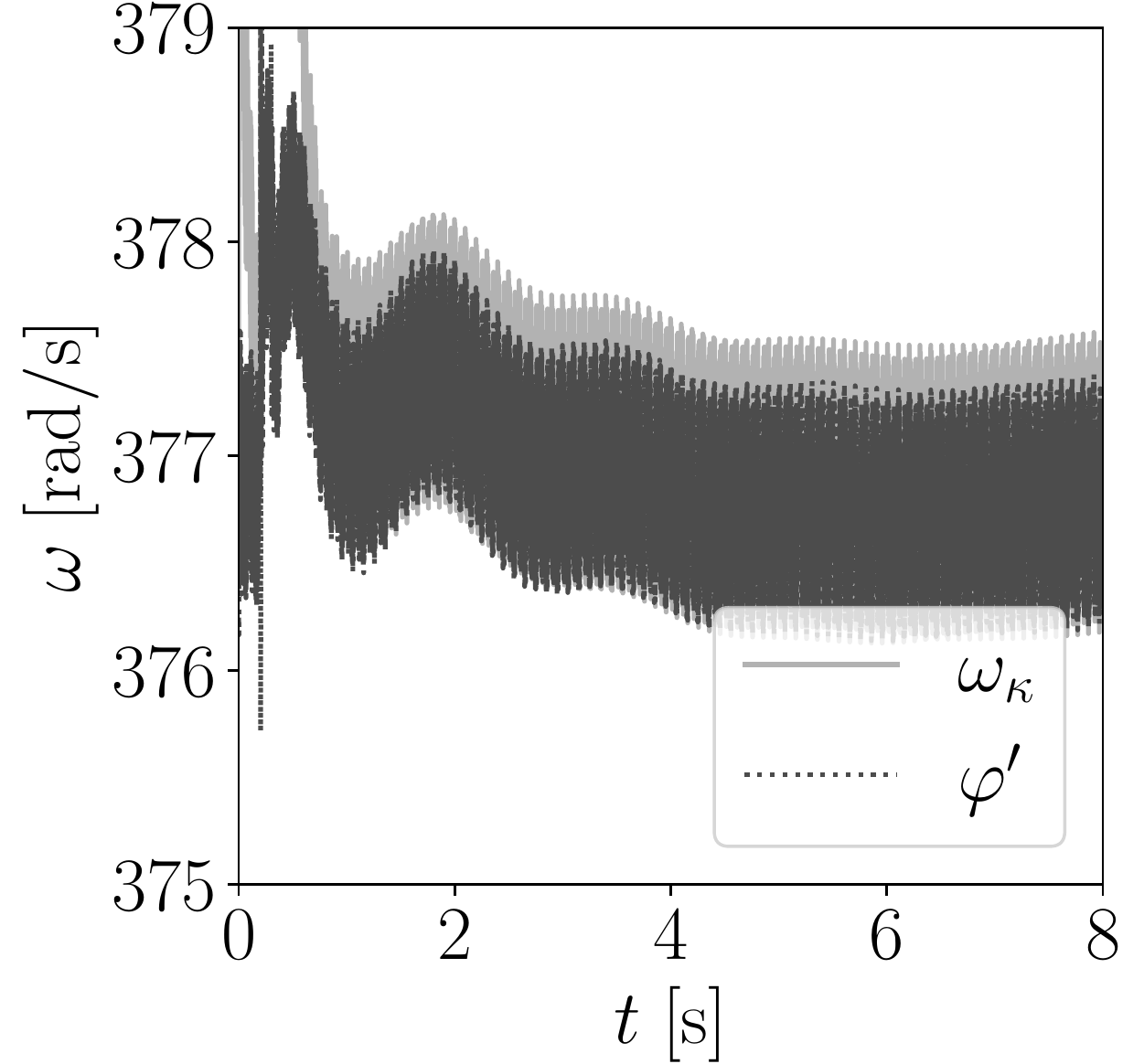}}}
    \subfigure[]{\resizebox{0.475\linewidth}{!}{\includegraphics{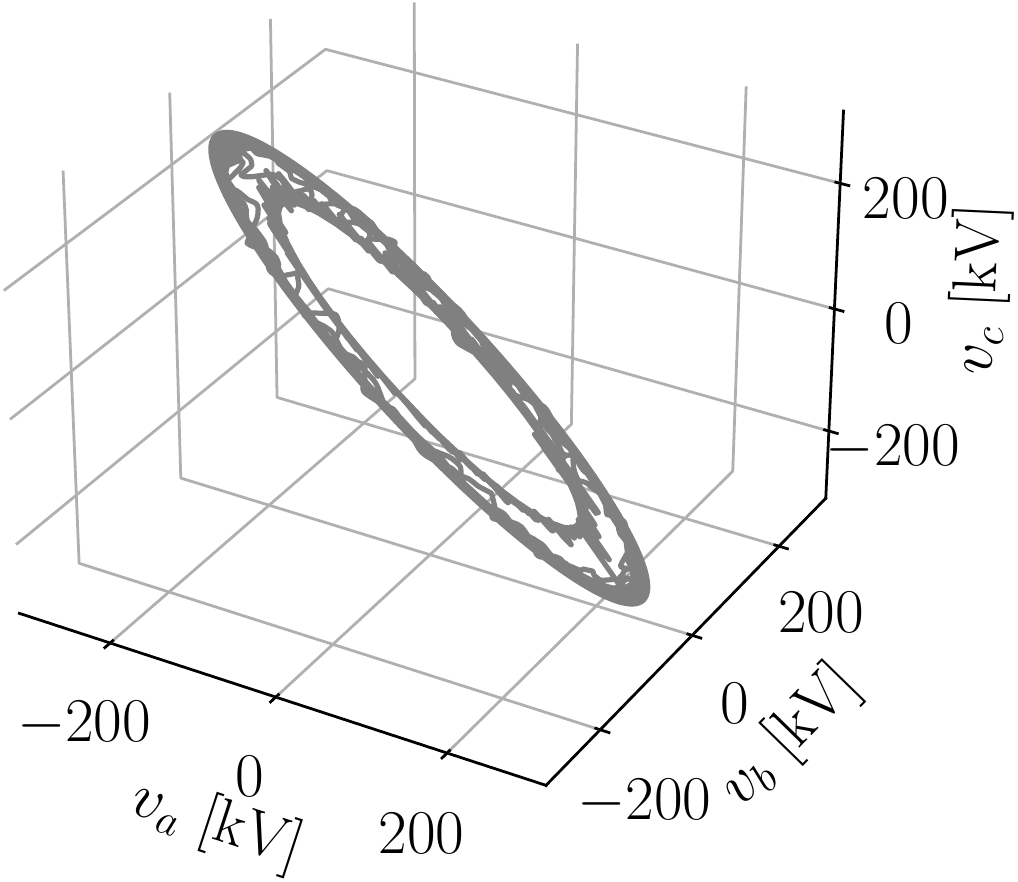}}}
    \subfigure[]{\resizebox{0.475\linewidth}{!}{\includegraphics{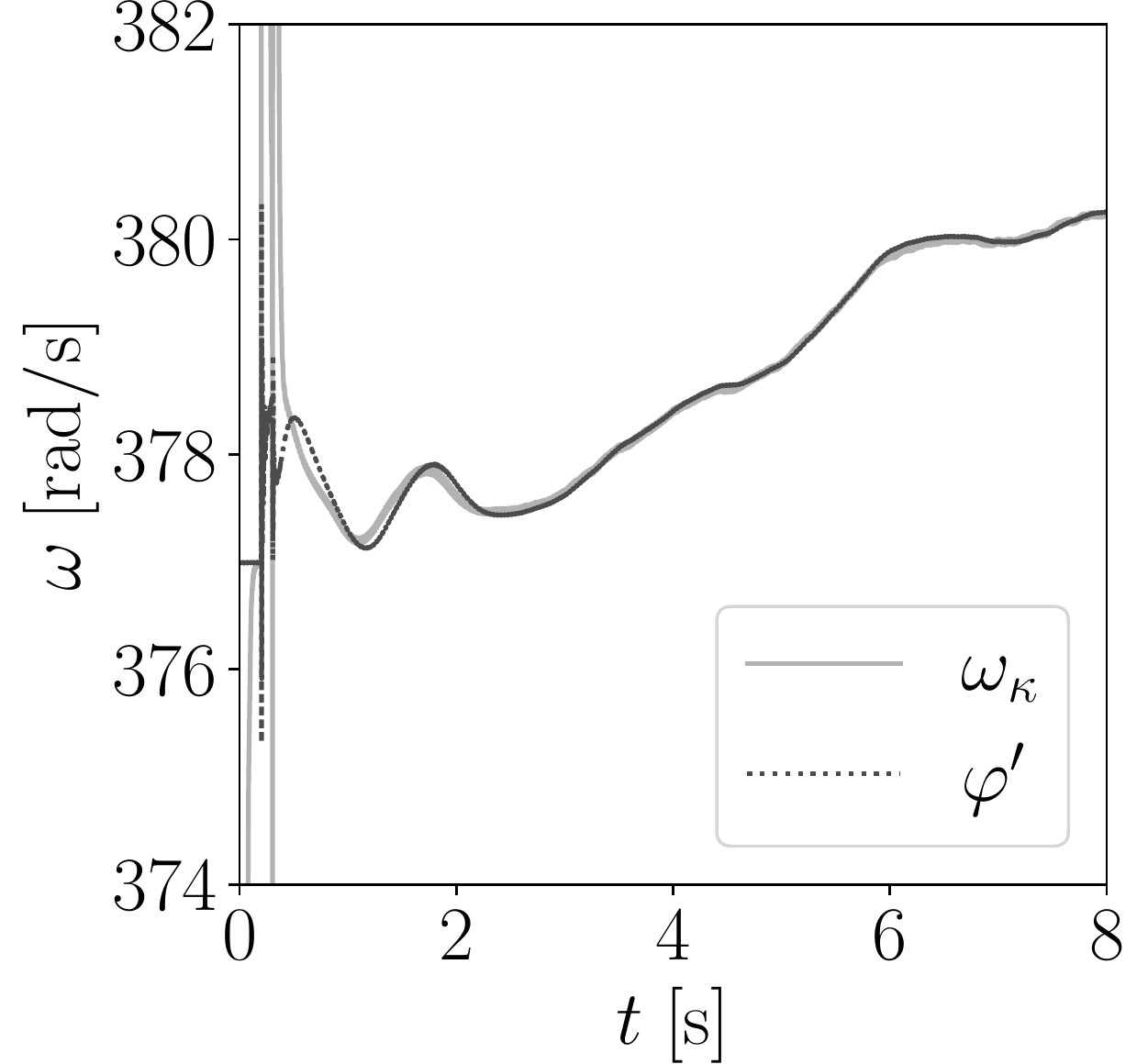}}}
    \caption{IEEE 39-bus system. (a, b) Base-case balanced system; (c, d) unbalanced system; (e, f) unbalanced harmonic current sources; and (g, h) balanced system with Gaussian noise. (a, c, e, g) Voltage at bus 26 in the space $(v_a, v_b, v_c)$; (b, d, f, h) estimated frequency.}
    \label{fig:39bus}
  \end{center}
\end{figure}

Figure \ref{fig:39bus} shows $\bfg v_{abc}(t)$, $\wk$ and $\varphi'(t)$ as obtained for four scenarios, as follows.

\begin{itemize}
\item \textit{Base-case balanced system}.  This is the benchmark system provided with DIgSILENT PowerFactory.
\item \textit{Unbalanced system}.  The power consumption of all 19 loads of the system is unbalanced, with imbalances ranging from $5$ to $10\%$ on one of the phases.
\item \textit{Unbalanced harmonic current source}. Unbalanced $5$-th and $7$-th harmonic current sources are added to bus 26.
\item \textit{Balanced system with noise}. A Gaussian noise is added to the measurement voltage input of all excitation systems of the synchronous machines.
\end{itemize}

Before the fault, the three phases are balanced and thus, the corresponding part of the curve is circular and lies in a plane.  The same holds after the fault clearance but the voltage converges to a circle that is different from that of the initial steady-state.  During the fault, the voltages are not perfectly balanced and symmetrical, which gives rise to the non-circular and non-planar sections observed in the left column of Fig.~\ref{fig:39bus}.

Regarding the estimation of the azimuthal frequency, we have adopted a simple numerical procedure.  First we have estimated the time derivatives using a numerical central derivative, i.e., the derivative of the voltage sample $v_i$ at time $t_i$ is approximated with:
\begin{equation*}
  v'_i \approx \frac{v_{i+1} - v_{i-1}}{2 h} \, ,
\end{equation*}
where $h$ is the time interval of the somapling rate and $v_{i+1}$ and $v_{i-1}$ are the voltage samples at $t_i + h$ and $t_i - h$, respectively.
Then we have calculated the azimuthal frequency using equation \eqref{eq:wk} and finally used a properly tuned lead-lag filter to remove high frequency noise.  The best results have been obtained using $h = 0.1$ ms.  For bigger $h$ the estimation becomes inaccurate and for smaller $h$ the effect of noise becomes dominant.  In turn, the estimation of the azimuthal (and, similarly, of the torsional frequency) presents same challenges and same tradeoffs as the estimation of the instantaneous frequency through PLLs.  
Overall, and as expected, there is a very good match between $\wk$ and $\varphi'(t)$, also for the scenarios with unbalanced voltages and harmonic contents.

\section{Conclusions}
\label{sec:conclusions}

The paper elaborates on the concept of frequency and discusses a geometric approach that allows defining a common framework for time- and frequency-domain approaches.  With this framework the well-known paradoxes of the \ac{if} can be explained in terms of curves and of coordinate transformations.  A variety of examples illustrate the proposed framework as well as the conditions for which the \ac{if} matches the expected shape and behavior of the frequency of a signal.  A case study based on the IEEE 39-bus system shows that \acp{pll} closely match the proposed geometric approach.

The geometrical approach appears promising and paves the way to a variety of future developments.  Reference \cite{freqfrenet} is our first work that utilizes the geometrical framework and focuses on circuit analysis.  We believe that this framework can be also exploited for practical applications, i.e., to design better controller and improve the dynamic performance of power systems, in particular the control of non-synchronous devices in low-inertia networks.  We are currently working in this direction.

\vfill

\end{document}